\documentclass[11pt,a4paper]{article}
\usepackage{amssymb,amsmath,graphicx,subcaption,hyperref,cite}
\usepackage{epstopdf}
\usepackage{tabularx}
\usepackage[utf8]{inputenc}
\usepackage[english]{babel}
\usepackage{multirow}
\begin{document}
\begin{center}{\Large {\bf Theoretical studies on switching of magnetisation in thin film}}\end{center}

\vskip 1cm

\begin{center} {\it {Moumita Naskar$^{1,2,a}$ and Muktish Acharyya$^{1,b}$ }}\end{center}

\vskip 0.5cm

\begin{center} {$^1$Department of Physics, Presidency University,\\
	86/1 College Street, Kolkata-700073, India.\\
	$^a$E-mail:naskar.moumita18@gmail.com\\
	$^b$E-mail: muktish.physics@presiuniv.ac.in\\}\end{center}

\vskip 1cm

\noindent {\bf Abstract:} 
In the present chapter, we focus on the switching of magnetisation, or the metastable lifetime of a ferromagnetic system. In this regard, particularly the Ising model and the Blume-Capel model, have been simulated in the presence of an externally applied magnetic field by the Monte-Carlo simulation technique based on the Metropolis algorithm. Magnetisation switching is found to be faster in the presence of disorder, modelled here by a quenched random field. The strength of the random field is observed to play a similar role to that played by temperature. Becker-D\"oring theory of classical nucleation (originally proposed for the spin-1/2 Ising system) has been verified in the random field Ising model. However, a stronger random field affects the nucleation regime. In a cubic Ising lattice, surface reversal time is found to be different from the bulk reversal time. That distinct behaviour of the surface in contrast to the bulk has been studied here by introducing a relative interfacial interaction strength ($R$). Depending on $R$, temperature, and applied field, a competitive switching of magnetisation of surface and bulk is noticed. The effect of anisotropy ($D$) on the metastable lifetime has been investigated. We report a linear dependency of the mean macroscopic reversal time on a suitably defined microscopic reversal time. The saturated magnetisation $M_f$, after the reversal, is noticed to be strongly dependent on $D$. $M_f$, $D$, and $h$ (field) are found to follow a proposed scaling relation. Finally, Becker-D\"oring theory as well as Avrami's law are verified in spin-$s$ Ising and Blume-Capel models. The switching time depends on the number of accessible spin states.


\vskip 2 cm

\noindent $^2$Present address: Institute of Mathematical Sciences,CIT Campus,\\ Tharamani, Chennai, Tamil Nadu 600113, India.
\newpage

\vskip 2cm

\noindent {\large \bf Objectives:}\\

\begin{itemize}
    \item How does disorder present into a system affect the switching of magnetisation?
    \item Comparative study of reversal of surface and bulk.
    \item Effects of magneto crystalline anisotropy on switching of magnetisation.
    \item Switching of magnetisation in spin-$s$ Ising and Blume-Capel  system 
\end{itemize}


\vskip 2 cm

\tableofcontents


\newpage


\section {Introduction}

\vskip 0.5cm


In modern technology as well as in our daily life, magnetic thin films are used in a wide variety of devices due to their enormous applications. Magnetic storage media (Piramanayagam and Chong, 2011; Daniel et al., 1998) are one such essential devices that play crucial role to store information in the form of tiny magnetic grains. By switching of magnetisation it is meant to drive the system's magnetisation in an opposite direction compared to the initial direction of magnetisation by applying an external magnetic field. Thin films have drawn great attention from the researchers due to their controllable divergent properties compared to the bulk material as a consequence of reduced coordination number, reduced symmetry, etc. Experimental studies of magnetisation reversal in thin films were started almost seven decades ago and it is still ongoing with immense interest in order to upgrade storage capacity, durability etc of the devices. We are particularly concerned about how fast or how slow the grains respond to the applied magnetic field.

For the faster recording or accessing of the data, the tiny magnetic grains are expected to respond quickly to the external field. At the same time, we should be aware of the stability of data against any kind of effective noises (thermal or magnetic field) (Vogel et al., 2006) for better longevity of the devices. So, for practical purposes, a compromise between these two cases is extremely important so that the switching time of the magnetisation can be tuned to the demand of the technological world. After all, it seems very useful if some theoretical knowledge guides us at the beginning of some real experiments. 

%
In the context of switching of magnetisation, the phenomenological Becker-D\"oring theory (Becker and D\"oring, 1935) is much appealing which nicely presents the reversal time (or so-called nucleation time) as a function of temperature and the magnitude of the applied magnetic field. Later, the prediction of this phenomenological theory was verified by the Monte Carlo simulation where the growth of droplets can be studied as phase ordering kinetics (Puri, 1999). The relaxation of Ising ferromagnet after a sudden reversal of applied magnetic field is also studied (Binder and M\"uller-Krumbhaar, 1974). The rate of nucleation of crystalline solids in a solid-melt system was explored in an important historical study (Grant and Gunton, 1985). The dependence of metastable lifetimes on the applied magnetic field and the system size was investigated extensively in kinetic Ising ferromagnet (Rikvold et al., 1994). Extensive simulational research on nucleation in different dimensions has been done using heat-bath dynamics, and a good consistency between the numerical results and the theoretical predictions of Becker-D\"{o}ring has been reported (Acharyya and Stauffer, 1998). The investigations of the thermally activated magnetisation switching of small ferromagnetic particles, involving coherent rotational motion and precessional motion, driven by an external magnetic field have been carried out in Hinzke and Nowak, 1999, 2002. Domain dynamics of magnetic films with perpendicular anisotropy have been reported (Nowak et al., 1997). The rates of growth and decay of the clusters of different sizes have been studied (Vehkam\"aki and Ford, 1999) as functions of external field and temperature. 
 
Asymmetric reversal modes in ferromagnetic/ antiferromagnetic multilayers were also studied (Beckmann et al., 2003). The distribution of nucleation times, in the system showing Brownian-type dynamics, has been described by classical nucleation theory (Brendel et al., 2003). 
.
The heat-assisted magnetisation reversal in ultrathin films for ultra-high-density information recording has been investigated (Deskins et al., 2011). In a recent paper (Acharyya, 2014), nucleation time was observed to increase in the presence of a magnetic field spreading over the space in time as compared to that in a static field. The linear reversal mechanism in FePt grains has been simulated using atomistic spin dynamics, parameterized from ab-initio  calculation (Ellis and Chantrell, 2015). Very recently, the magnetisation reversal in Ising ferromagnet driven by a spatially graded field (Dhar and Acharyya, 2016) along with the presence of a thermal gradient and a marginal competition (between field gradient and thermal gradient) has been reported (Dutta et al., 2018).


Metastability and nucleation in the Spin-1 Blume-Capel (BC) (Blume, 1966; Capel, 1966, 1967a, 1967b) ferromagnet were studied and found the different mechanisms of transition (Cirillo and Olivieri, 1996). They report an abrupt change in the mechanism of transition (from a metastable state to the stable state) when crossing a certain value of field (two times the chemical potential). The metastability in the BC model with distributed anisotropy was studied (Yamamoto and Park, 2013) using different dynamics. Extensive results regarding the critical properties of the general spin $s > 1$ Blume-Capel model can be found in the works by Plascak and collaborators (Plascak et al., 1993; Plascak and Landau, 2003). For the particular spin-1 case, we refer the reader to Refs. (Fytas et al., 2018, Vatansever et al., 2020). Anyway, apart from the metastability, the influence of magneto-crystalline anisotropy on other phenomena like dynamical phase transition, universality class, critical properties, etc. are explored to a great extent using the Blume-Capel model (Costabile et al., 2012; Silva et al., 2006; Gulpinar et al., 2012;, Yeomans and Fisher, 1981). 

The mixed spin ($s=1,1/2$) Blume-Capel model was investigated (Selke and Oitmaa, 2010) by Monte Carlo simulation and the absence of a tricritical point was noticed in two dimensions. The magnetic properties of mixed-integer and half-integer spins in a Blume-Capel model were studied (Masrour et al., 2017) by Monte Carlo simulation.

\vskip 0.5 cm
\noindent $\Box$ {\large \bf Becker-D\"oring theory of classical nucleation and metastable lifetime}
\vskip 0.1 cm

In order to study the switching of magnetisation or the reversal of magnetisation theoretically, one should focus on the metastability possessed by the system in presence of a magnetic field. The switching time indicates the lifetime of the metastable state which appears in presence of the applied field. Since the Becker-D\"oring theory is the most reliable theory to explain the dynamics of metastability in the magnetic system, it is good to start with a very short discussion on it. How does a ferromagnetic system respond to a weak magnetic field if applied to the system below critical temperature $T_C$, in the opposite direction to that of the net initial magnetisation? Initially, the system enters into a metastable state which eventually decays to the stable equilibrium state only if the applied field is sufficient to overcome the energy barrier of the metastable state (Gunton and Droz, 1983; Vehkam\"aki, 2006). How does the metastable state appear? And, how does it decay eventually? Historically, the classical nucleation theory aimed to answer all those questions by introducing the dynamical and statistical characteristics of the nucleation process (Becker and D\"oring, 1935; Gunton and Droz, 1983). Specifically, the dynamics of metastability were analyzed by R. Becker and W. D\"{o}ring in 1935.

For a ferromagnetic (spin-1/2 Ising) system, in presence of a negative external field, classical nucleation theory assumes that the small droplets of down spins are dispersed in the background of up spins. The number of such droplets of down spins is assumed to follow Boltzmann distribution,
$n_l = Ne^{-\beta E_l}$
where $\beta=1/k_B T$ and $E_l$ is the free energy of formation of a droplet of size $l$ ($l$ number of down spins) and $N$ is the normalization factor. The classical assumption is that $E_l$ comes from the contribution of bulk energy and surface energy. In presence of a negative magnetic field, an energy barrier ($E_c$) results from the competitive behaviour of these two terms. 

The Becker-D\"{o}ring theory (Gunton and Droz, 1983; Becker and D\"oring, 1935) explains the behaviour of metastability by the kinetics of cluster (droplets of spins) formation. The basic assumption of this theory is that the time evolution of the number of droplets is only due to an evaporation-condensation mechanism in which a droplet of size $l$ loses or gains a single spin. Any type of coagulation or other kinds of interactions is neglected here. The ultimate prediction of this phenomenological theory is that the nucleation rate intimately depends on the energy barrier $E_c$ which is dependent on the field.
\begin{equation}\label{eq:6}
I=I_0 e^{-E_c/k_BT}
\end{equation}
where $I_0$ is the rate prefactor. For a weak applied field, reversal occurs through the growth of a single supercritical droplet. The nucleation time or the metastable lifetime in the nucleation regime (NR) is simply inversely proportional to the nucleation rate $I$ derived by Becker-D\"{o}ring theory, 

\begin{equation}
\tau_{(nr)} \sim I^{-1} \sim exp\Bigg(\frac{K_d\sigma^d}
{k_BTh^{d-1}}\Bigg)
\end{equation}

where $K_d$ is the dimension dependent term arising on differntiating the free energy term with respect to the droplet size. In contrast, for stronger applied field, the reversal occurs through the coalescence of many critical droplets. 

In that coalescence regime (CR), the reversal time is obtained by

\begin{equation}
\tau_{(cr)} \sim I^{-\frac{1}{d+1}} \sim exp\Bigg(\frac{K_d\sigma^d}
{k_BT(d+1)h^{d-1}}\Bigg)
\end{equation}

So if we plot the logarithm of the reversal time against $1/h$, it would be a fair straight line. In addition, the slope of the straight line in the coalescence regime will be smaller compared to the slope observed in the nucleation regime.
\begin{figure}[htbp]
\centering
	\begin{subfigure}[b]{0.49\textwidth}
	\centering
	\includegraphics[width=\textwidth,angle=0]{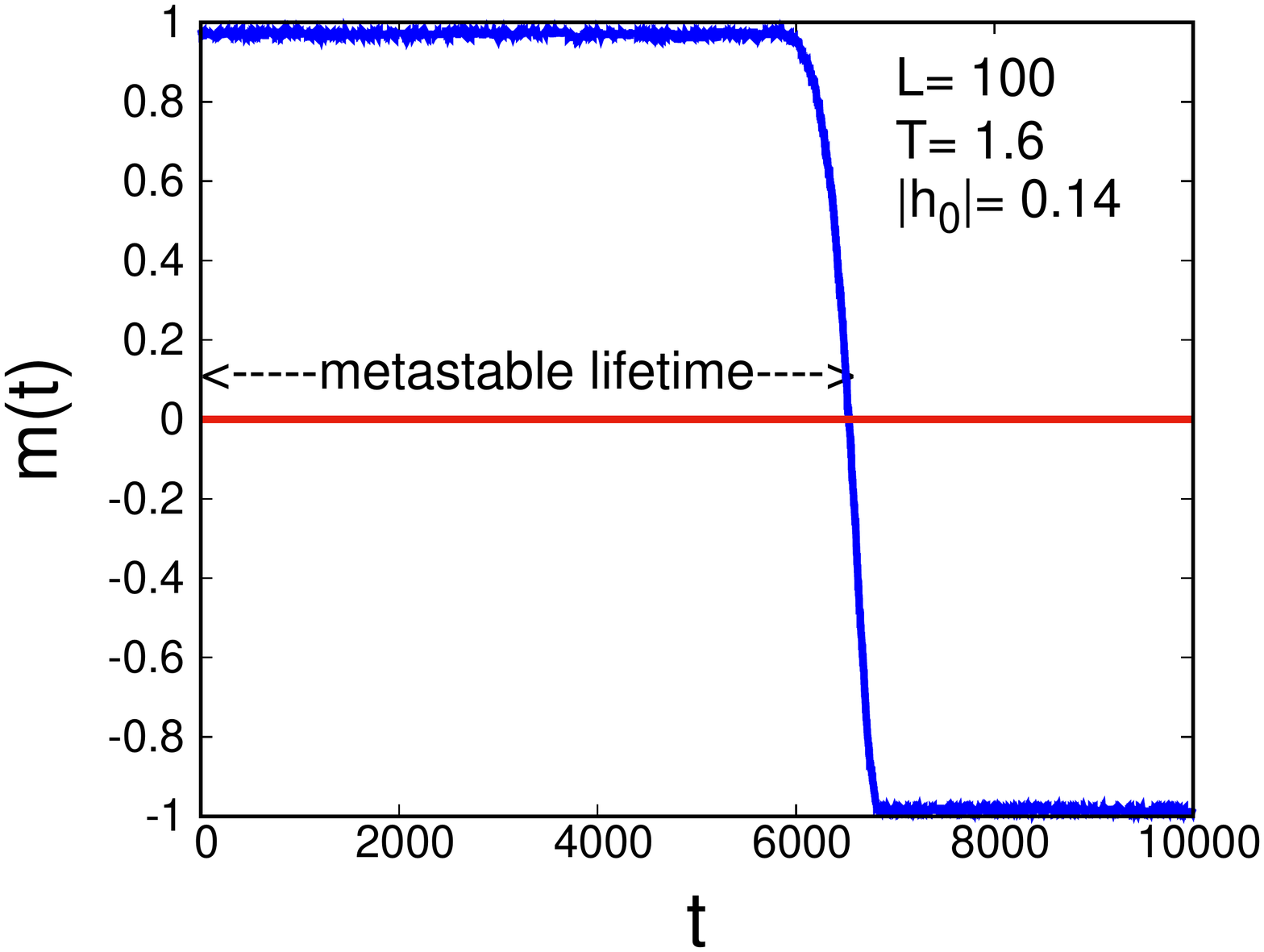}
	\subcaption{}
	\end{subfigure}
	\begin{subfigure}[b]{0.495\textwidth}
	\centering
	\includegraphics[angle=0,width=\textwidth]{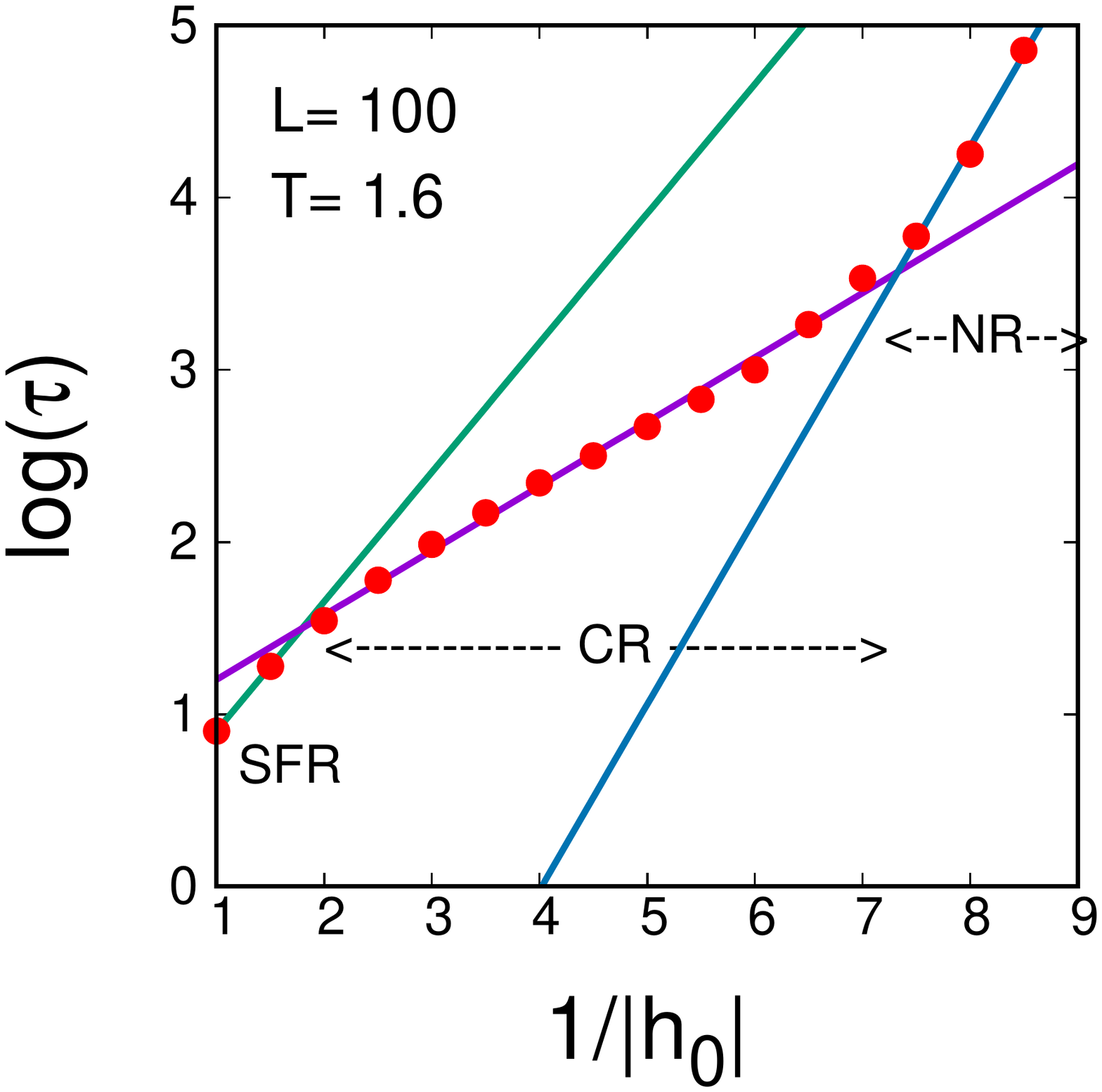}
	\subcaption{}
	\end{subfigure}
\caption{\footnotesize (a) A typical decay of metastable state of Ising system in the presence of external field $h_0=-0.14$. (b) Logarithm of mean metastable lifetime with inverse of the applied field. Temperature is set to $T=1.6(0.7T_c)$ and the lattice size is $L=100$ for both the plots.  Naskar M. and Acharyya M., 2020, Physica A
{\bf 551}, 124583}

\label{meta_state}
\end{figure}
A typical decay of metastable state of a single two dimensional Ising sample has been illustrated in Fig-\ref{meta_state}a by studying the variation of magnetisation with time. The results are obtained here by the Monte-Carlo simulation technique based on Metropolis dynamics (Binder and Heermann, 1992). Clearly, the system first enters into the metastable state (a flat portion with positive magnetisation). Later, it decays to a stable state ($m(t) \simeq -1$). It may be notified here that, we have defined the lifetime of metastable state ($\tau$) as the minimum time required to achieve negative magnetisation ($m(t)\simeq 0$) from a completely ordered state. The metastable lifetime is also referred to as the reversal time or the switching time of magnetisation. 
Since we are involved in statistical analysis of the results, it is always recommended to check the results over different samples. So in the upcoming sections, the discussion will be based on the behaviour of the `mean or average reversal time' obtained by the arithmetical average of the reversal times of different samples.


Becker-D\"oring results are already well verified in spin-1/2 Ising system by Monte Carlo simulation. Here it is illustrated once for convenience \ref{meta_state}b. Logarithmic mean metastable lifetime (calculated over 5000 samples) is plotted with $1/|h_0|$ at a fixed temperature ($T=1.6$) well below the critical temperature. Three distinct regimes of different reversal mechanisms (strong field regime (SFR), coalescence or multi-droplet regime (CR) and nucleation or single-droplet regime (NR)) are clearly identified. Smaller slope in the coalescence regime, as predicted by Becker-D\"oring theory, is also well verified here.

Fundamentally, the switching time depends on the system's temperature and applied magnetic field which is well predicted by Becker-D\"oring theory. But, how does it vary with the presence of disorder, anisotropy as well as with the number of spin states? How is it modulated by some spatial variation of field and anisotropy? Furthermore, does the surface show distinct behaviour of reversal compared to the bulk? The following discussion will shed some light on those matters.


\section {Model and computer simulation scheme}
\label{model}

In the present chapter, the whole discussion will be centered around the computer simulational studies of metastable lifetime of ferromagnets in two discrete classical spin models namely the Ising model and the Blume-Capel model. The simulation method involves particularly the Monte Carlo simulation technique based on Metropolis algorithm. By considering the discrete nature of the spins and neglecting other quantum effects, these classical spin models have been highly popular and successful in explaining the thermally activated phase transition like the ferromagnetic-paramagnetic transition in magnetic systems. Ising model is a good prototype to study the magnetic properties of a system. Blume-Capel model is the simplest spin model which gives some insights of the effect of magneto crystalline anisotropy. 

The Hamiltonian of the general spin-$s$ Blume-Capel model (Blume, 1966; Capel, 1966), where $s$ assumes integer or half-integer values of the spin, is represented by,

\begin{equation}
\mathcal{H} = - \frac{1}{s^2}\;J \sum_{\langle i,j\rangle} s_i^z s_j^z + \frac{1}{s^2}\; D \sum_i (s_i^z)^2 - \frac{1}{s}\; h \sum_i s_i^z,
\end{equation}\\
where $s$ is the total spin. $s_i^z$ denotes the $z$-component of the spin which takes values from $-s$ to $+s$ through unit steps. For example, for the $s = 5/2$ spin system, the values of $s_i^z$ are well known $s_i^z  = {5/2,3/2,1/2,-1/2,-3/2,-5/2}$. Now it should be clearly notified that, we have considered here the normalized values $s_i^z/s$ of z-component of the spin. For example, for the $s = 5/2$ spin system, $s_i^z/s= \sigma_i^z= {1,3/5,1/5,-1/5,-3/5,-1}$. Definitely, for $s=1/2$ and $D=0$, the Hamiltonian will recover the Hamiltonian for the original Ising system where spin can take two values $s_i^z/s=+1,-1$ only. Whereas, for $s=1$ and non-zero value of $D$, it will recover the Hamiltonian of the original Blume-Capel system where spin takes the values $s_i^z/s=+1,0,-1$. 
The first term in the Hamiltonian describes the exchange interaction between the nearest neighbour spins. $J$ ($ > 0$) is the uniform ferromagnetic exchange interaction strength between the nearest neighbour spins only. Ferromagnetic behaviour is implemented here by considering the positive $J$. The second term models here the effect of single-ion anisotropy (or, crystal-field coupling) $D$. The third term indicates the interaction of individual spin with the applied magnetic field ($h$). Both $D$ and $h$ have been measured in units of $J$ and the temperature has been used in the unit of $J/k_B$.

Let us discuss now the numerical protocol we have used in the following investigations. Started with a perfect ordered state ($\sigma_{i}^{z} = +1 \; \forall i$), the lattice is updated by random updating scheme. The simulation has been dealt with square or cubic lattice. Either periodic or open boundary conditions are applied according to the topic of study. A site (i-th say) has been chosen randomly. The present value of $s_i^z$ at that chosen site is $s_i^z(initial)$. The updated value may be any of the values between $s_i^z=-s $ to $s$. The final trial state of $s_i^z$ is chosen randomly from any of these $2s+1$ values with equal probability. Let this test value be labeled as $s_i^z(final)$. The probability of $s_i^z$, to assume the final value $s_i^z(final)$ from its initial value $s_i^z(initial)$, is determined by Metropolis transition probability, 
\begin{equation}
P(s_i^z(initial) \to s_i^z(final)) = {\rm Min} {\Big [}1,{\rm exp} {\big (} {- \frac{\Delta \mathcal{H}}{k T}} {\big )} {\Big ]},
\end{equation}
\noindent where $\Delta \mathcal{H}$ is the change in energy (calculated from equation-1) due to the change in the value of $s_i^z$, from $s_i^z(initial)$ to $s_i^z(final)$. $k$ is the Boltzmann constant and $T$ is the temperature of the system. The temperature of the system is measured in the unit of $J/k$. For simplicity, we set $J=1$ and $k=1$ throughout the simulational study. The acceptance of the final value $s_i^z(final)$ is determined by comparing a random number with the Metropolis transition probability. The test move is accepted only when the random number (uniformly distributed in the range [0,1]) is less than or equal to $P(s_i^z(initial) \to s_i^z(final))$. In this way, the total $L^d$ number of randomly chosen spins (random updating scheme) are updated, where $d$ is the dimension of the system. $L^d$ number of such random updates constitutes one Monte Carlo Step per Spin (MCSS) which acts as the unit of time in the problem.

The instantaneous magnetisation of the system is determined by
\begin{equation}
m(t) =\frac{1}{L^d} \sum_{i}^{L^d} \sigma_i^z  
\end{equation}

\vskip 1cm

\section {Discussion based on simulational results:}

\vskip 0.5 cm


\subsection{Switching of magnetisation in a disordered system}

Most of the popular order-disorder phase transitions in nature are primarily driven by thermal fluctuation. Recently, people find interest in investigating the influence of any other kind of quenched disorder on such transitions. For such studies, the random field Ising model (RFIM) is a good prototype which was a remarkable discovery proposed by Imry and Ma, 1975. 

How does disorder affect the reversal of magnetisation? This is a pertinent question and should be addressed in the research of the magnetisation switching phenomena because disorder, defects etc are an unavoidable phenomena which often remains in natural material as well as appears during synthesizing new materials. In this context, we found some relevant earlier reports. Various types of heterogeneous nucleation has been studied in nanoscale ferromagnetic grains using Ising model (Kolesik et al., 1997). Heterogeneous nucleation has also been studied in two dimensional Ising model where the impurities were placed on a line of fixed points (Scheifele et al., 2013). Here, we will briefly cover the influence of quenched random field on metastable lifetime using random field Ising model (Naskar and Acharyya, 2020).

The Hamiltonian of such a randomly disordered Ising ferromagnetic system is represented by,
\begin{equation}
\mathcal{H}=-J\sum_{<i,j>}\sigma_i^z \sigma_j^z - \sum_{i}h_i \sigma_i^z
\end{equation}
Terms are already discussed in details in the previous section \ref{model}. Here, we will talk about the form of field only. $h_i$ is the random field which is the resultant of an externally applied uniform field ($h_0$) and a quenched random field ($h_r$) i.e. $h_i=h_0+h_r$. It is worth mentioning that, $h_r$ is a quenched random field which is considered here to model the effect of disorder. Moreover, the mean of $h_r$ is set to zero ($<h_r>=0$) so that $<h_i>=h_0+<h_r>=h_0$. So the mean of total field $h_i$ remains $h_0$ which helps to draw a comparison between the system's behaviour in presence of a uniform field and that in presence of a random field. 
\begin{figure}[ht!]
\begin{center}
\begin{subfigure}{0.495\textwidth}
	\includegraphics[width=1.05\textwidth,angle=0]{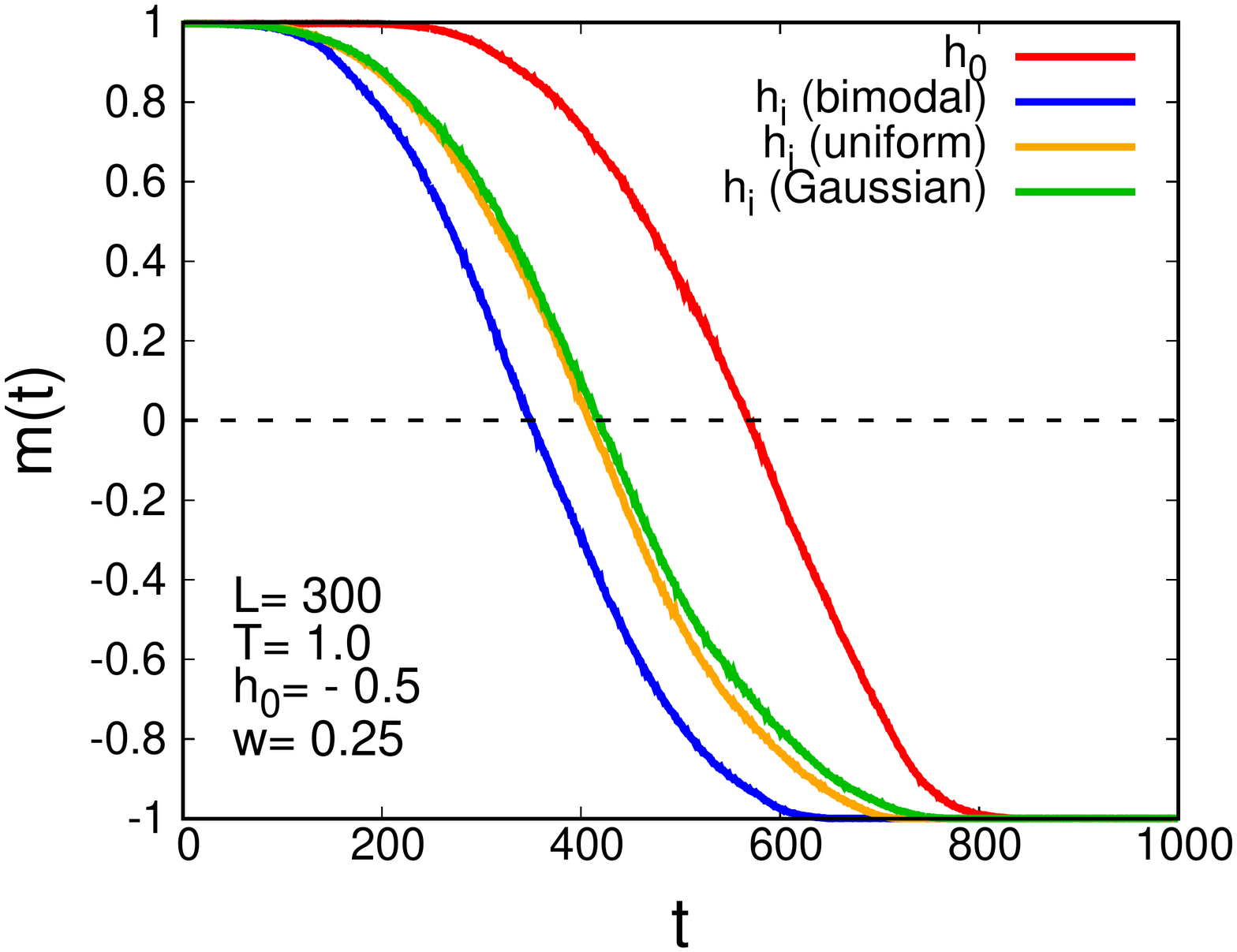}
	\subcaption{}
\end{subfigure}
\begin{subfigure}{0.495\textwidth}
	\includegraphics[width=1.1\textwidth,angle=0]{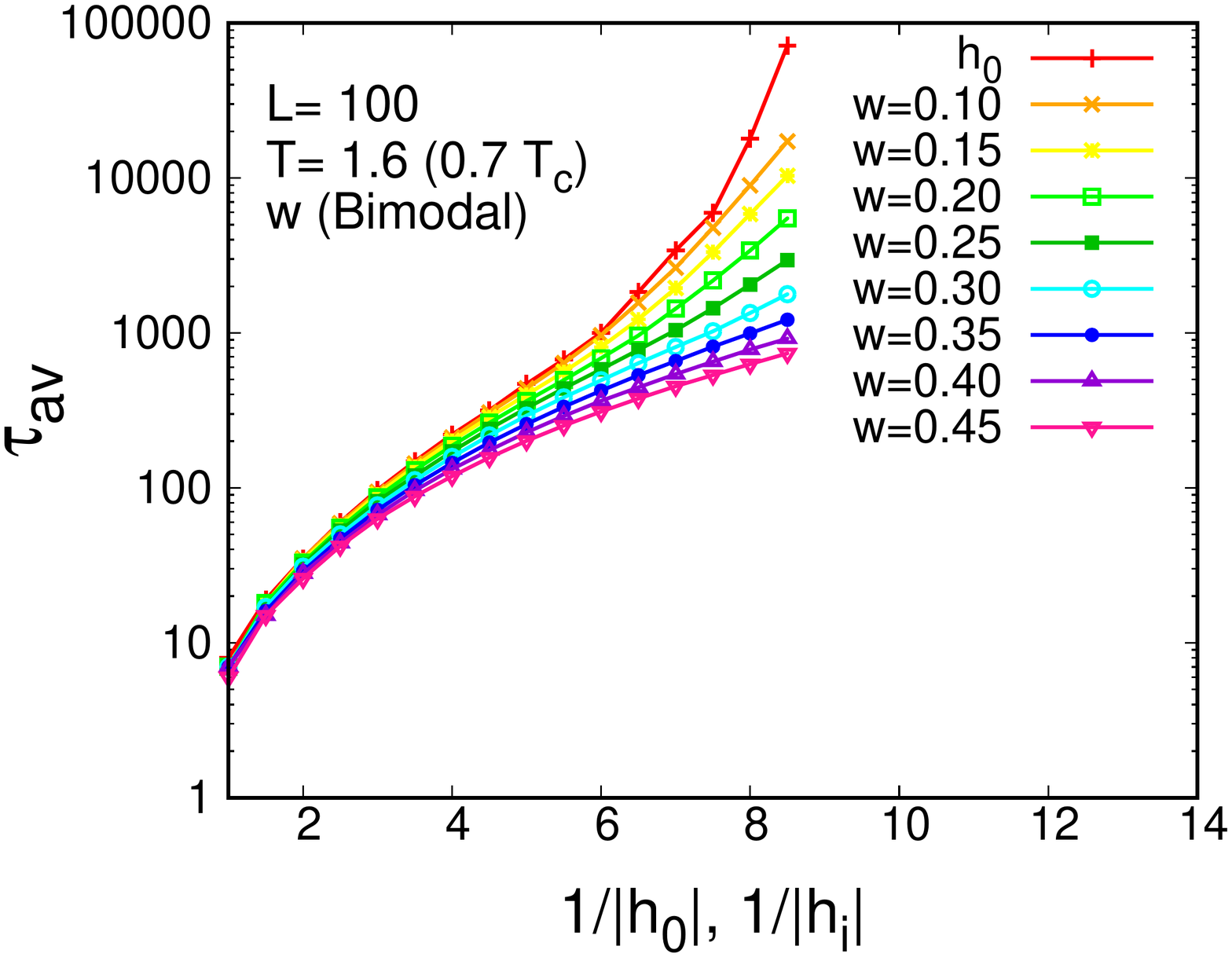}
	\subcaption{}
\end{subfigure}
\caption{\footnotesize (a) Variation of magnetisation with time at temperature $T=1.0$ in presence of $h_0$ and $h_i$ (for three distributions of $h_r$). The uniform field is set to $h_0=-0.5$ and the width of random field is set to $w=0.25$. Lattice size is $L=300$. (b) Variation of the logarithmic mean reversal time ($\tau_{av}$) with the inverse of field ($h_0$ and bimodal $h_i$ having different $w$) for lattice size $L=100$ at temperature $T=1.6$.  Naskar M. and Acharyya M., 2020, Physica A,
{\bf 551}, 124583}
\label{2magtime}
\end{center}
\end{figure}

Three different kinds of distributions of random field ($h_r$) have been used. (a) Bimodal distribution, 
$	P_b(h_r)=0.5\delta(h_r-w)+0.5\delta(h_r+w)$
which implies that, approximately 50 $\%$ of lattice sites experience the field $h_r=+w$ whereas the rest of the lattice sites experience the field $h_r=-w$ which are definitely randomly placed into the lattice sites. (b) Uniform distribution,
$	P_u(h_r)=\frac{1}{2w}$
where $h_r$ is uniformly distributed between $-w$ to $+w$ throughout the lattice in random manner. (c) Gaussian distribution, 
$	P_n(h_r)=\frac{1}{\sqrt{2\pi\sigma^2}}e^{-\frac{h_r^2}{2\sigma^2}}$
with standard deviation $\sigma=2w$ and $\mu=0$. Box-Muller algorithm has been used to generate normally distributed random numbers. 

In the above distributions, we have denoted the ``width'' or ``strength'' of the random field by `$w$' which indicates that the random field ($h_r$) are distributed from $h_r=-w$ to $h_r=w$. That means the value of the total random field varies from $h_i= h_0-w$ to $h_i= h_0+w$. To avoid any confusion, let me mention clearly that, the actual width of the distribution of random field ($h_i$) is $2w$ but for the sake of simplicity, we have denoted it simply as $w$.
The ferromagnetic Ising square lattice of size $L$ has been simulated with periodic boundary conditions applied in both directions. The system is updated by random updating scheme using Metropolis algorithm as mentioned earlier in the previous section.

First of all, the evolution of the magnetisation $m(t)$ with time $t$ has been studied for a single sample (Fig-\ref{2magtime}a), in the presence of a uniform field ($h_0$) as well as in the presence of three different kinds of random field ($h_i$) of strength $w=0.25$. Let me clarify once more the meaning of `$h_i$ of strength $w=0.25$'. That means, $h_r$, which is randomly distributed between $-0.25$ and $+0.25$ following three different probability distributions, is added to $h_0$. Fig-\ref{2magtime}a depicts that the metastable lifetime decreases in presence of any kind of random fields. Bimodal distribution is found to be more effective comparatively (maybe the reflection of its discrete symmetry unlike the other two distributions). In addition, the metastable lifetime varies with the nature of the distribution of $h_i$ though each distribution of $h_i$ has mean $h_0$. 
\begin{figure}[h!]
  \begin{subfigure}[b]{0.5\textwidth}
    \includegraphics[width=0.33\textwidth,angle=-90]{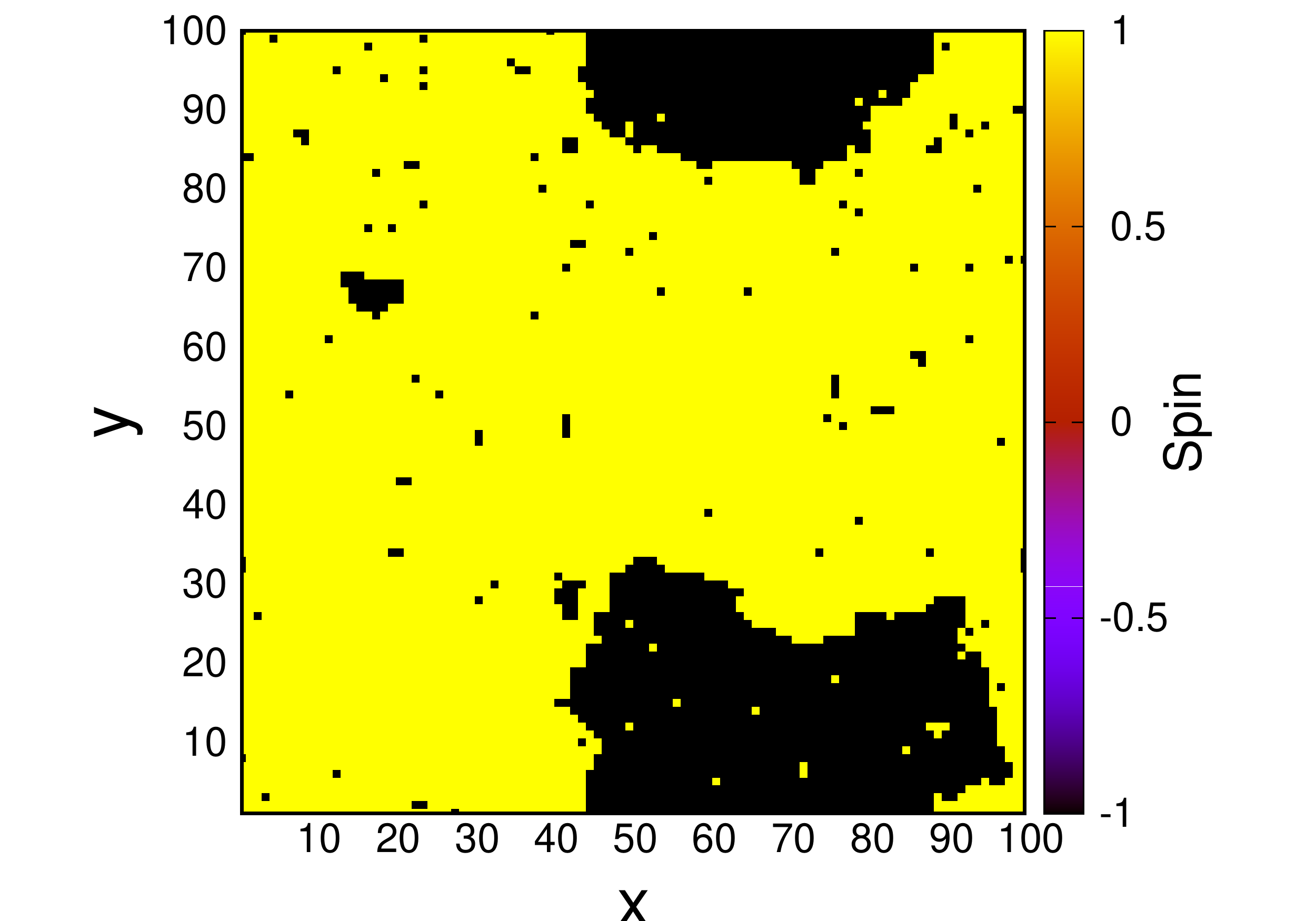}
    \includegraphics[width=0.33\textwidth,angle=-90]{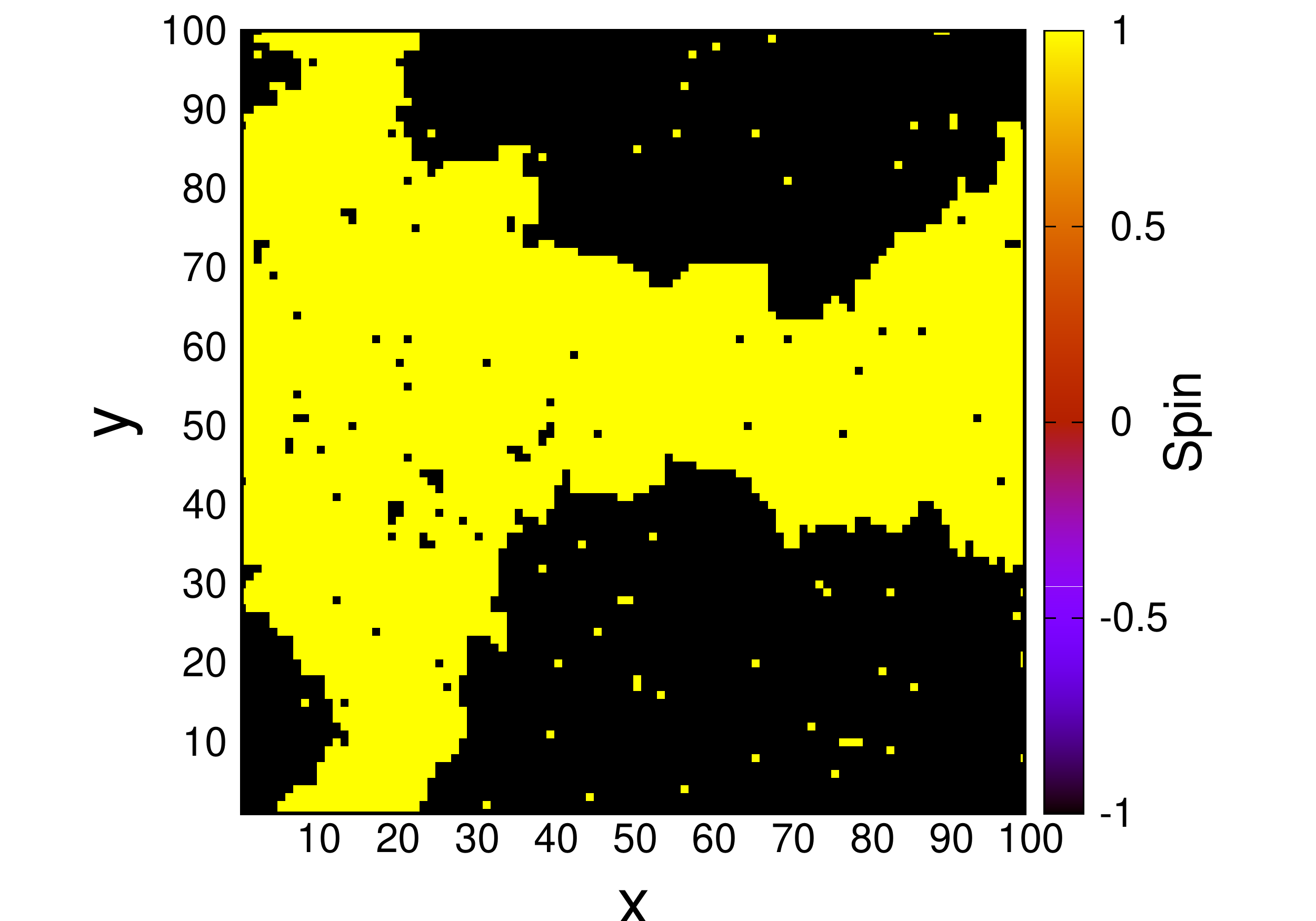}
    \subcaption{$w=0.3$}
  \end{subfigure}
  \begin{subfigure}[b]{0.5\textwidth}
    \includegraphics[width=0.33\textwidth,angle=-90]{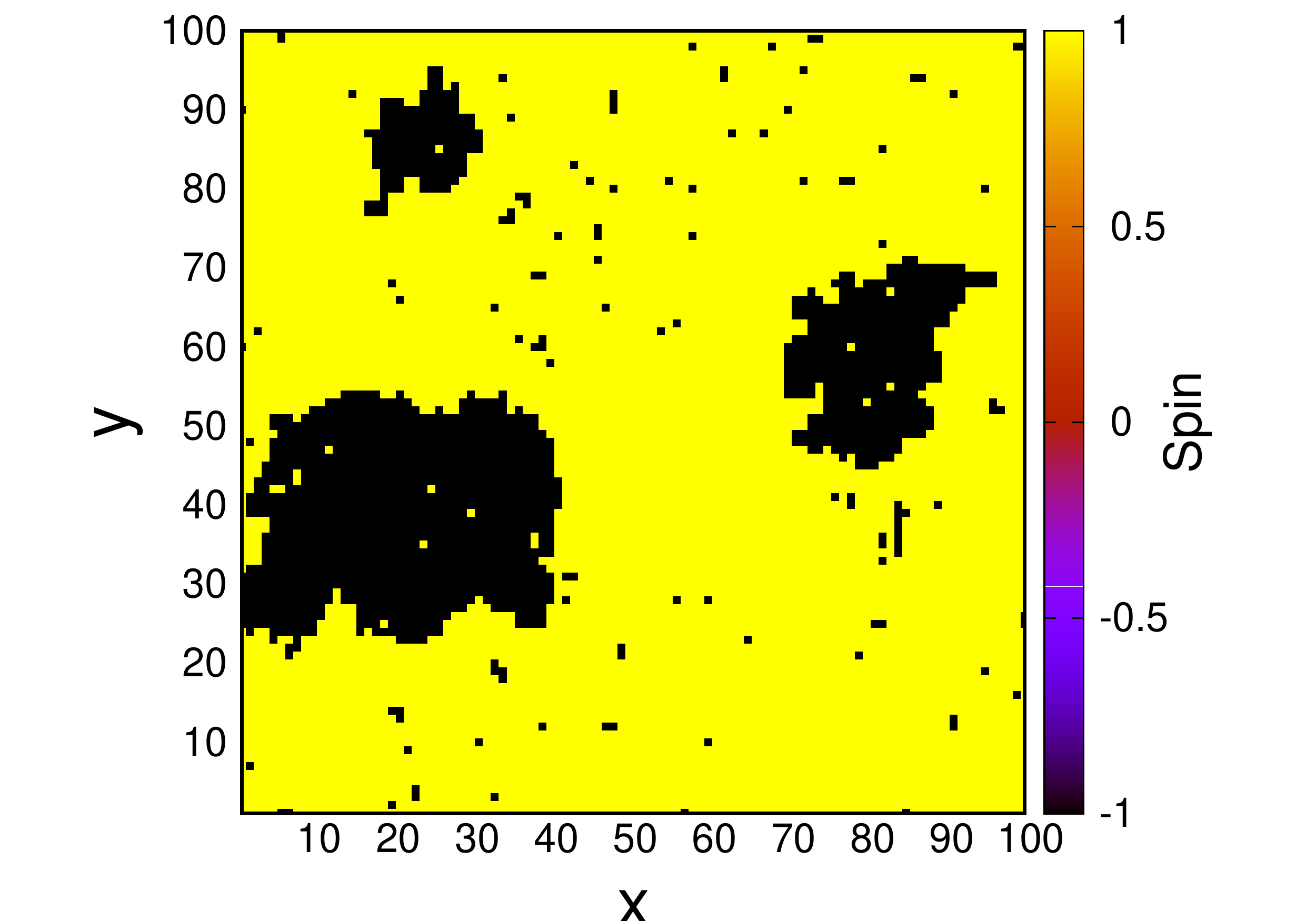}
    \includegraphics[width=0.33\textwidth,angle=-90]{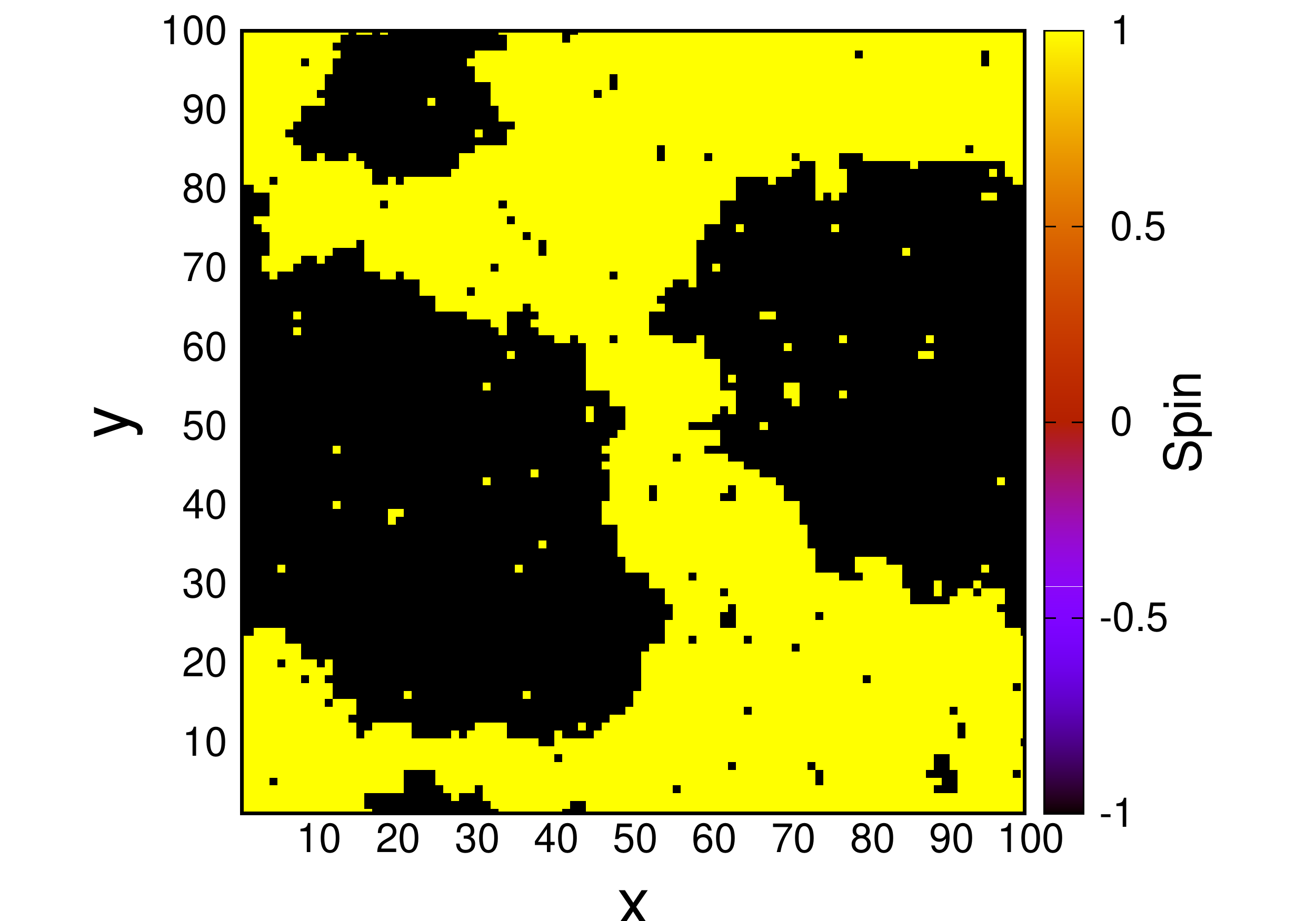}
    \subcaption{$w=0.35$}
  \end{subfigure}
\caption{\footnotesize Image plots of spin configuration at two different Monte-Carlo single steps, in the presence of $h_i$ (bimodal distribution) (a) of width $w=0.3$ and $h_0=-0.125$, left: $t=2300$ (before reversal), right: $t=2546$ (at reversal). (b) of width $w=0.35$ and $h_0=-0.125$, left: $t=500$ (before reversal), right: $t=711$ (at reversal). Naskar M. and Acharyya M., 2020, Physica A,
{\bf 551}, 124583.}
\label{2snap_bi} 
\end{figure}

In order to compare with the Becker-D\"{o}ring analysis, we have studied (Fig-\ref{2magtime}b) the variation of mean reversal time `$\tau_{av}$' with the inverse of applied magnetic field (bimodal random distribution) by keeping the temperature well below the critical temperature $T=1.6(0.7 T_c)$. Mean reversal time is calculated from 5000 different samples. 

Fig-\ref{2magtime}b confirms that the mean metastable lifetime decreases in presence of bimodal random field compared to that in presence of uniform field only (red curve). More interestingly, as the strength of the random field is increased, the strong field and the coalescence regimes are not affected significantly but the weak field regime i.e. the nucleation regime becomes unclear slowly.
Snapshots of spin configurations near the reversal time in the nucleation regime ($\langle h_r \rangle = h_0= -0.125$) for two different strengths of bimodal random field $w=0.3 \& 0.35$, reveal that the system is no more in nucleation regime for $w=0.35$ (Fig-\ref{2snap_bi}). So we can say that there must be a limiting value of the width of the random field between $w=0.3$ and $w=0.35$ beyond which the nucleation regime no more exists. It is an interesting observation, which was not reported before. 

The possible reasons can be analyzed in the following way. Generally, the system exhibits nucleation in presence of a very weak applied field. For a distribution of stronger random field (such that $w > |h_0|$), which exceeds the value of uniform field $h_0$, makes the field $h_i$ effectively stronger at some lattice sites. Suppose the system is in nucleation regime with $h_0= -0.125$ or $\frac{1}{|h_0|} = 8$. Now if we apply a bimodal random field ($h_r$) having width $w= 0.3$ i.e. $50\%$ field is $h_i= -0.125-0.3 = -0.425$ and $50\%$ field is $h_i= -0.125+0.3 = 0.175$. Obviously, the $50\%$ net positive field is totally unable to flip the spin. So the effective strength of the field would be $h_i= -0.425$ which is stronger compared to $h_0= -0.125$ but can affect only $50\%$ of sites approximately. As a consequence, up to a certain value of $w > |h_0|$, the system still remains in the nucleation regime. Beyond that critical value of $w$, the nucleation regime disappears completely and the system stays in the multi-droplet regime.  


Similar studies as above have been carried out in presence of uniform and Gaussian random fields also. The results for those two distributions, we would like to refer Naskar and Acharyya, 2020. In both the cases, unlike the bimodal case, we observed that the system remains in nucleation regime even in presence of

a random field of width $w=0.45$ which has been confirmed by taking the snapshots in the same fashion as described for the bimodal case. In order to explain the fact, we take the same example as bimodal distribution. Suppose the system is in the nucleation regime with $h_0= -0.125$. Now if we apply a uniform random field ($h_r$) having width $w= 0.3$ i.e. all the values of field between $h_i= -0.125-0.3 = -0.425$ and $h_i= -0.125+0.3 = 0.175$ will be distributed in equal proportion. Obviously, the lattice sites having the net field values from $h_i= 0$ to $h_i=0.175$ will not participate in the microscopic reversal mechanism. Rest of the sites, experiencing the negative values of the field distributed from $h_i= -0.425$ to $h_i=0$, will play an active role in the reversal process. Similar facts are expected for Gaussian distribution also. So clearly the uniform and Gaussian distribution are less effective (in affecting the nucleation regime) compared to the bimodal distribution.

\begin{figure}[h!]
\centering
  \begin{subfigure}[b]{0.495\textwidth}
  \centering
    \includegraphics[width=1.1\textwidth,angle=0]{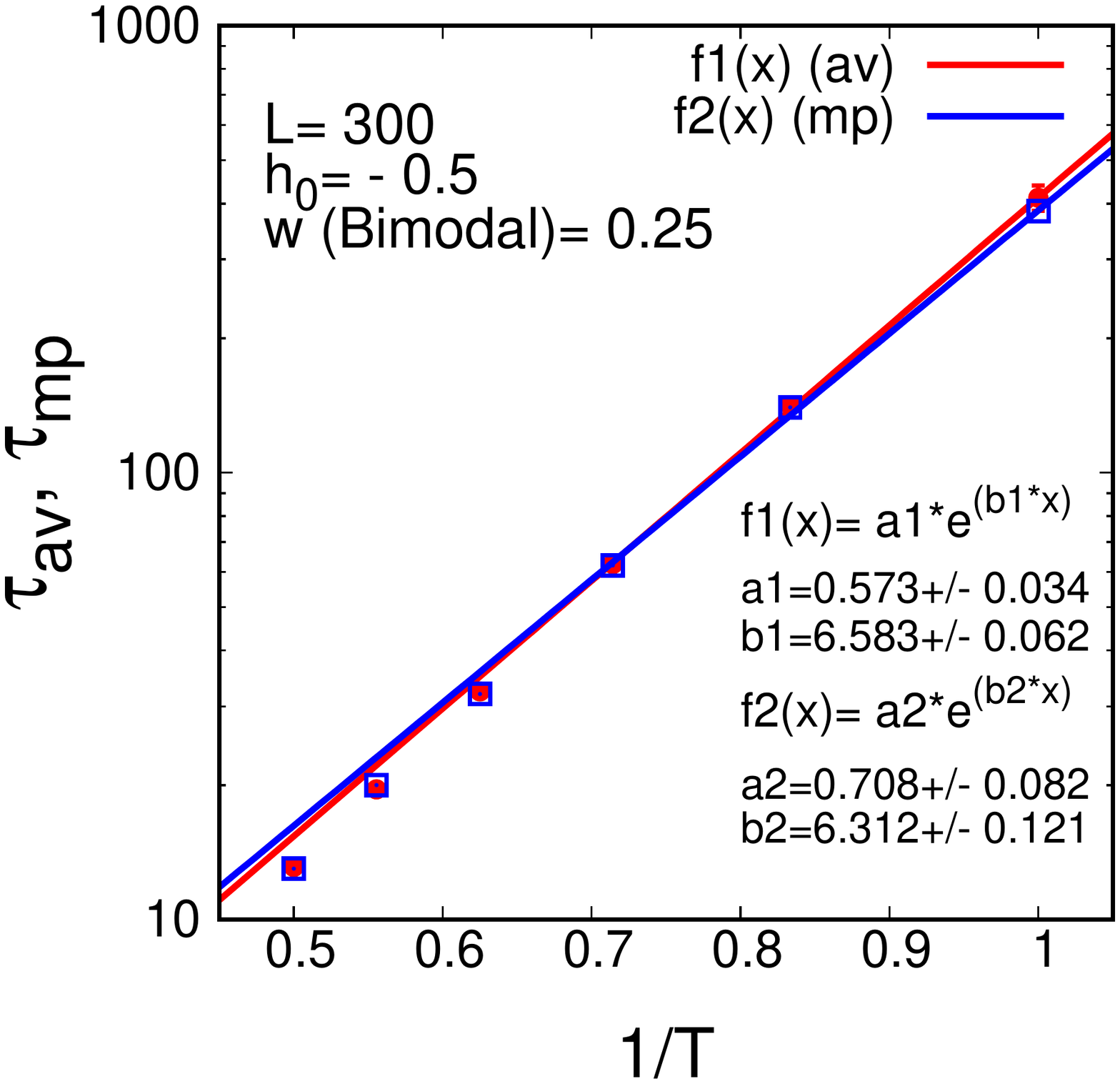}
    \subcaption{}
  \end{subfigure}
  \begin{subfigure}[b]{0.495\textwidth}
  \centering
   \includegraphics[width=1.1\textwidth,angle=0]{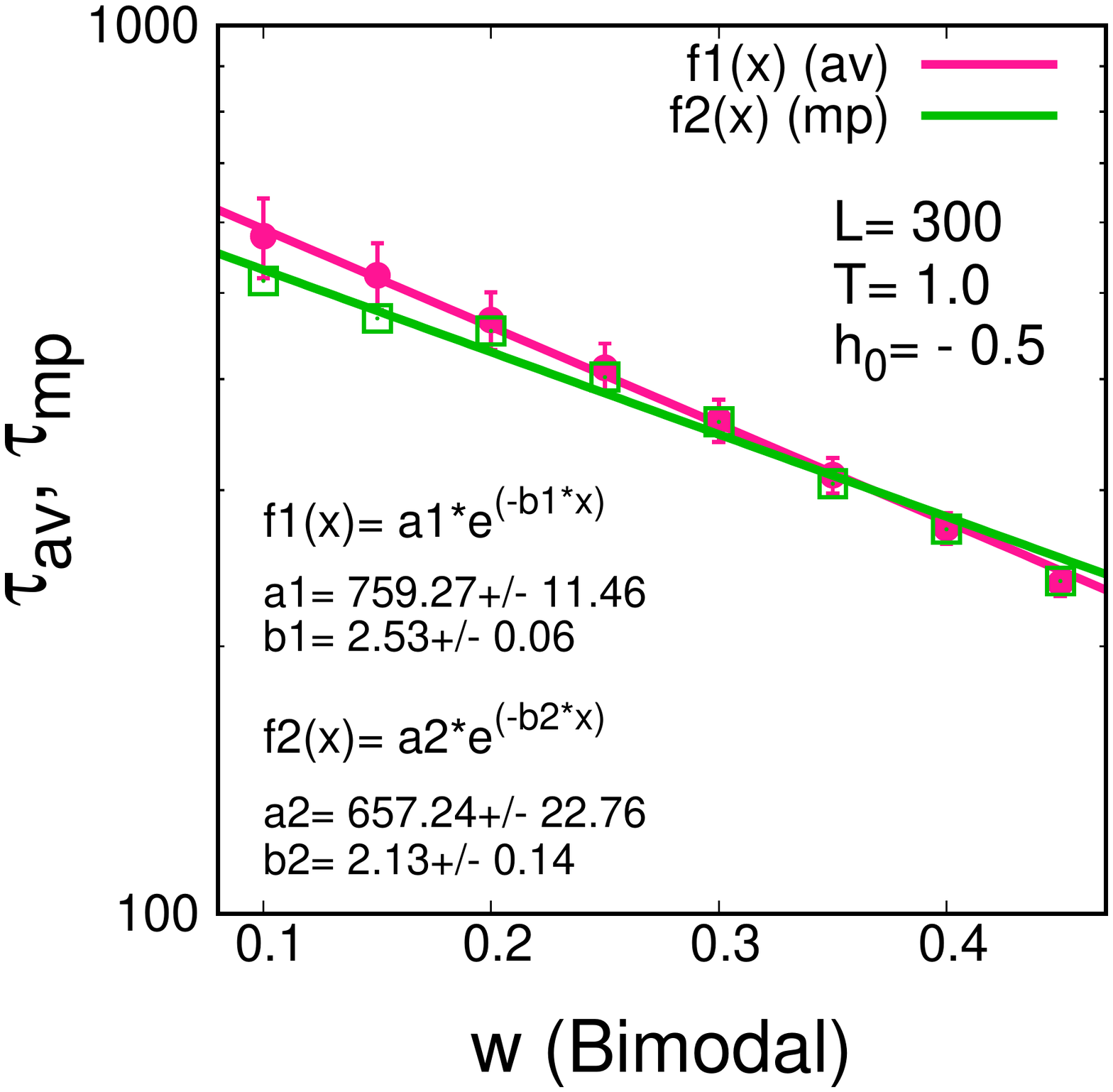}
    \subcaption{}
  \end{subfigure}

\caption{\footnotesize Variation of mean reversal time ($\tau_{av}$) and most probable reversal time ($\tau_{mp}$) with (a) the inverse of temperature in presence of the bimodal random field $h_i$. In each case, uniform field is $h_0=-0.5$ and the width of the random field is $w=0.25$ (b) the width of bimodal random field $h_i$. Uniform field is $h_0=-0.5$ and the temperature is $T=1.0$. Lattice size is $L=300$. Naskar M. and Acharyya M., 2020, Physica A,
{\bf 551}, 124583}
\label{2tau_temp-RF}
\end{figure}
\vskip 0.5 cm
\noindent $\Box$ {\large \bf Influence of temperature and random field:}

Variation of both the mean and most probable reversal time has been studied with the inverse of temperature in presence of three different distributions of the random field of the same strength $w=0.25$. We observed that (Fig-\ref{2tau_temp-RF}a) both the mean and the most probable reversal time increase exponentially with the inverse of temperature as predicted by Becker-D\"{o}ring theory in presence of the uniform field. Additionally, the error bar (which is the standard deviation of the reversal times of different samples here) of the reversal time increases with the decrease in temperature. 

In the same fashion the dependence of reversal times on the strength or width of the random field disorder has been examined (Fig-\ref{2tau_temp-RF}b). Reversal times decrease exponentially with the increase of $w$. And also the error bar (standard deviation of the reversal times of different samples) of the reversal time decreases with the increase in $w$. Interestingly, one can tune the metastable lifetime of magnetisation by varying temperature as well as the strength of random field. So it can be inferred in a qualitative sense that, the random field disorder $w$ is equivalently playing the role of temperature $T$.

\vskip 0.5cm


\subsection{Competitive reversal of magnetisation of surface and bulk}

Magnetism of surface has drawn much attention of the researchers due to its divergent behaviours compared to the bulk. First of all, let me highlight some very recent interesting observations briefly. In Park and Pleimling, 2012, the role of surfaces at nonequilibrium phase transitions has been elucidated using the Ising system in presence of an oscillating magnetic field. 
 
Surface phase diagram of the three-dimensional kinetic Ising model below the equilibrium critical point has been explored in presence of a periodically oscillating magnetic field (Tauscher and Peimling, 2014). 

Another study regarding the dynamic phase transition (DPT) of the kinetic Ising system has been accomplished very recently within the mean field approximation (Riego and Berger, 2015). Varying the surface exchange coupling strength, the amplitude of the externally applied oscillating field and its period, they explored some nonuniversal dynamic behavior of the layer-dependent magnetisation and the associated DPTs. 

Being inspired by those above studies involving significant effects of the surface, in particular, we opted to explore the behaviours of metastable lifetime of both the surface and bulk of the Ising system in presence of a negative applied field (Naskar and Acharyya, 2021b). The effect of surface can be modeled in different ways. Here we have followed a different approach by introducing an interfacial exchange interaction between the surface and core. The Hamiltonian is represented by,
\begin{equation}
	\mathcal{H}=-\sum_{<i,j>} J\: \sigma_i^z \sigma_j^z -\sum_{<l,k>} J\: \sigma \prime_l^z \sigma \prime_k^z -\sum_{<p,q>} J_f \: \sigma_p^z \sigma \prime_q^z - h \sum_{i}(\sigma_i^z+\sigma \prime_i^z)
\end{equation}	
where $\sigma_i^z$ (spin at i-th site of the core) and $\sigma \prime_l^z$
(spin at l-th site of the surface) are the Ising spins. The first term represents the contribution coming from the nearest neighbour interaction between the spin pairs within the core. The second term considers the interaction between pair of spins on the surface only. The third term captures the contribution coming from the interaction between the interfacial spins of core and surface. The last term represents the interaction of individual spins with the applied external magnetic field $h$. The nearest neighbour interaction strength between the core and the surface is taken as $J_f$ and all other types of (spin-spin) interactions are taken as $J$. Both $J$ and $J_f$ are ferromagnetic ($> 0$). For better realization we would like to refer the schematics of the lattice in Naskar and Acharyya, 2021c. It may be mentioned here that a relative interaction strength is defined as $R=J_f/J$ and $J$ is kept $J=1$ throughout the study for convenience.


Here, the surface consists of all the six outermost square layers of the cubic lattice. The system is kept in the open boundary conditions in all three directions. It may be worth mentioning here that such kind of interfacial interaction strength was used (Park and Pleimling, 2012; Tauscher and Pleimling, 2014; Riego and Berger, 2015) to study the surface critical behaviour of the non-equilibrium phase transition in driven kinetic Ising ferromagnets. It should be clarified that the whole system of $N=L^3$ number of spins is defined here as bulk. And the surface contains $N_s=L^3-(L-2)^3$ number of spins. For the convenience of discussion, let me denote the inner part of the bulk (that means excluding the surface) as the core. Notably, there exist three different categories of spins on the surface as far as the coordination number is concerned. Proportionally such corner and edge spins are so small compared to the other spins on the surface, we can neglect that matter.   

The total magnetisation of the bulk and the surface are determined by
\begin{equation}
m_b(t)=\frac{1}{N} \sum_{i}^{N} (\sigma_i^z + \sigma \prime_i^z)
\;\;\;\;\;\; \& \;\;\;\;\;\;
m_s(t)=\frac{1}{N_s} \sum_{ i } \sigma \prime_i^z 
\end{equation}
respectively, where $N=L^3$ is the total number of spins in the system. 

where $N_s= L^3-(L-2)^3$ is the total number of spins on the surface. 

\begin{figure}[htbp]
	\centering
\begin{subfigure}{0.495\textwidth}
	\includegraphics[angle=0,width=1.1\textwidth]{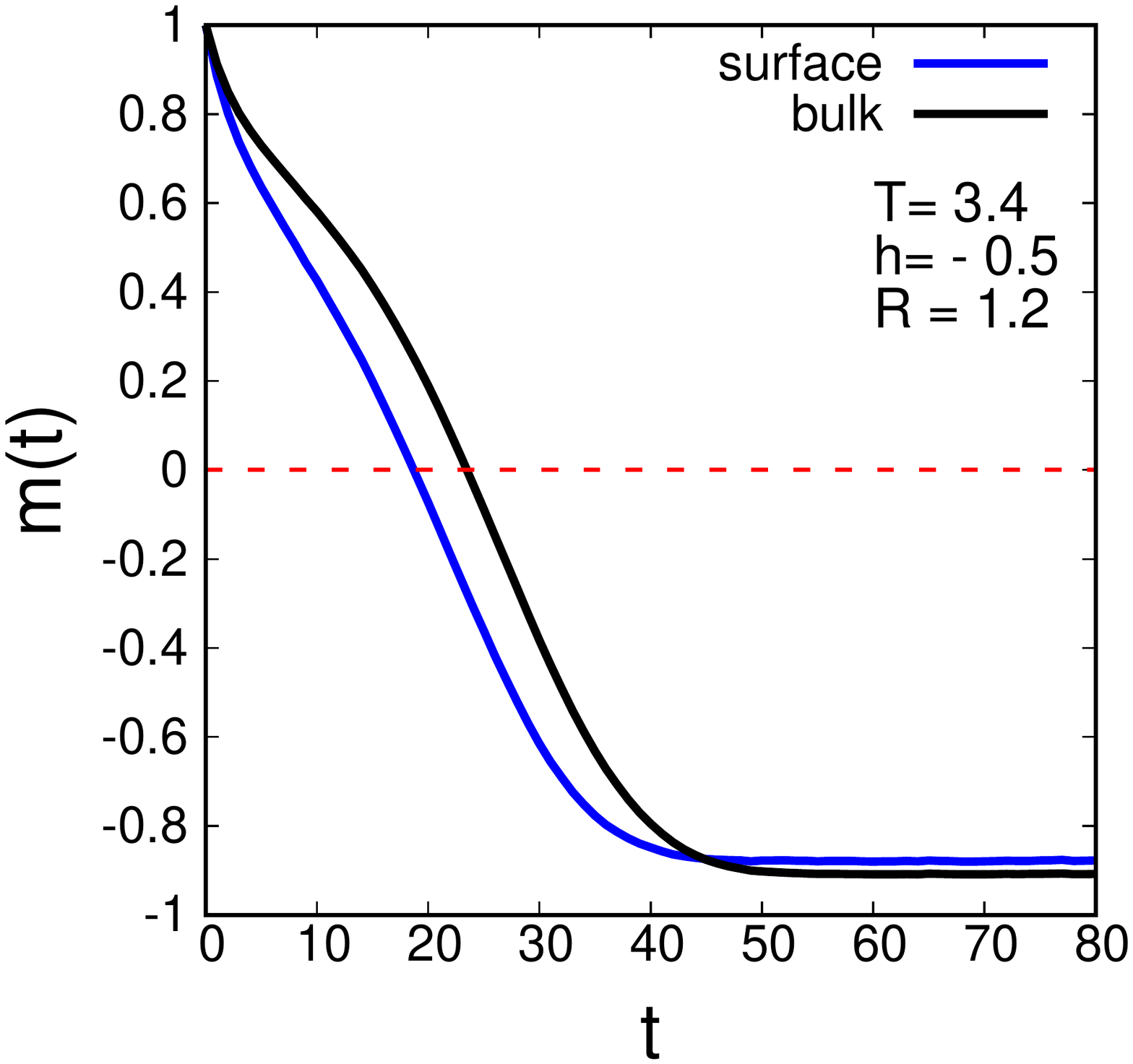}
	\subcaption{}
\end{subfigure}
\begin{subfigure}{0.495\textwidth}
	\includegraphics[angle=0,width=1.1\textwidth]{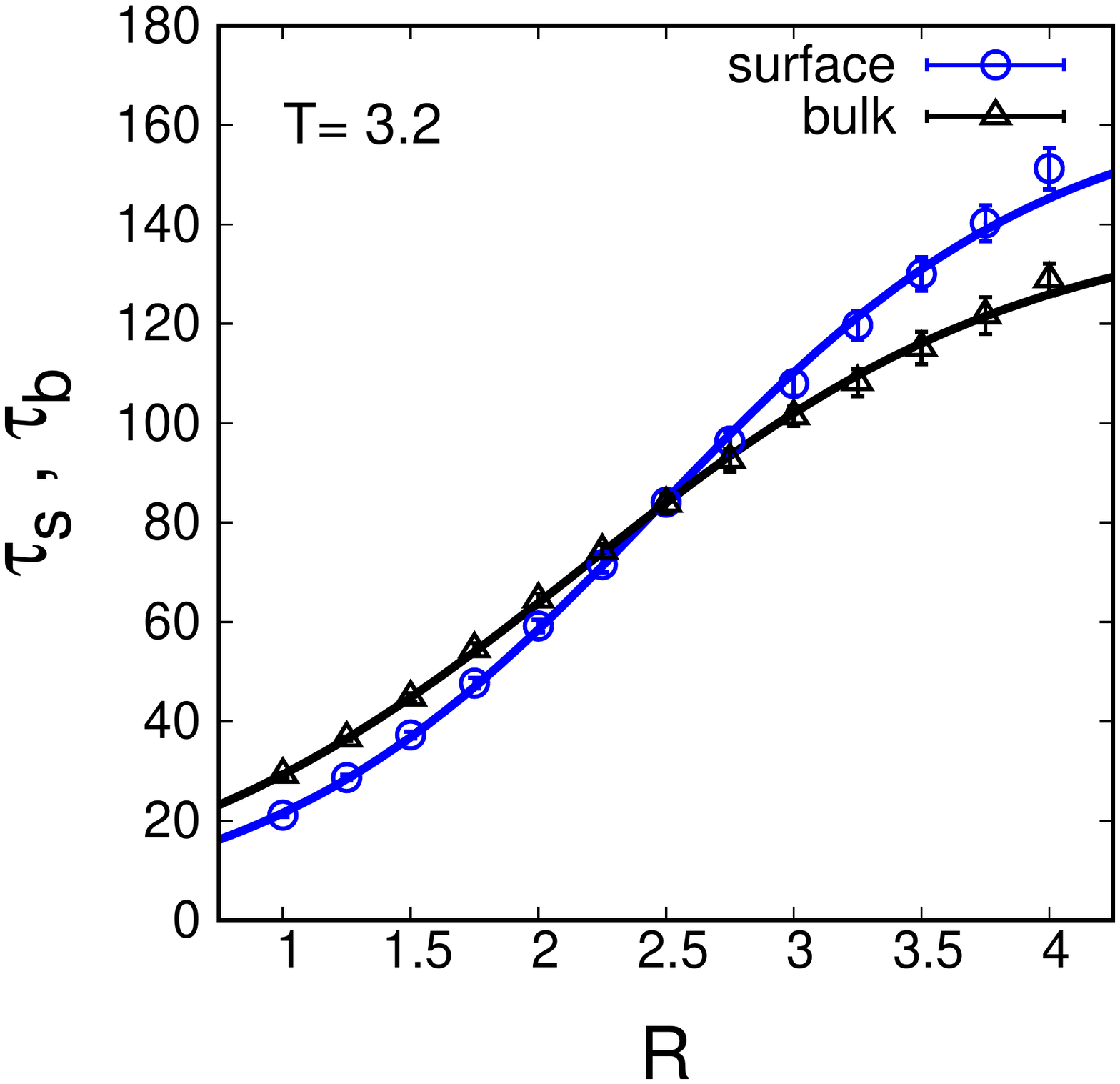}
	\subcaption{}
\end{subfigure}
\caption{\footnotesize (a) Variation of magnetisation $ m(t) $ with time $t$ of a single sample for $ R= J_f/J= 1.2 $. $ J=1.0$ always. Temperature is kept fixed at $T=3.4$ ($ \simeq 0.75 T_c$) (b) Mean reversal time of surface $\tau_s$ and bulk $\tau_b$ with the relative interaction strength $R$ at temperature $T=3.2$. Both the data for $\tau_s$ and $\tau_b$ fit to the function $f(x)= \frac{a}{1+e^{(b-x)/c}}$. In both (a) and (b) the applied field is $h= -0.5$ and the system size is $L=32$.
 Naskar M. and Acharyya M., 2021, Eur. Phys. J. B, {\bf 94}, 140.}
\label{3magtime}
\end{figure}

\vskip 0.5 cm
\noindent $\Box$ {\large \bf Metastable lifetime of surface and bulk:}
Starting from an initial state where all the spins are up (+1), the evolution of magnetisation with time has been studied for a single sample (Fig-\ref{3magtime}a) separately 
for surface and bulk in presence of a negative field $h= -0.5$ at a particular value of relative interaction strengths $R=J_f/J$. Temperature is kept well below the critical temperature of a three dimensional Ising system (Ferrenberg and Landau, 1991). Temperature and the applied field is chosen in such a way that the system is in multi-droplet or coalescence regime. The reversal time of the surface is found to be significantly different from that of the bulk. 

We also present here the variation of mean surface reversal time $\tau_s$ and mean bulk reversal time $\tau_b$, determined by averaging over the reversal times obtained for 1000 different samples, with $R$ (Fig-\ref{3magtime}b). In an obvious manner, both the $\tau_s$ and $\tau_b$ increase with the increase of $R$ as the stronger coupling helps to increase the longevity of metastable state. In the lower regime of $R$ ($J_f \sim J$), the magnetisation of the surface gets reversed faster compared to the bulk since fewer nearest neighbours enhance the probability of flipping of a spin on the surface. Now, if the $J_f$ is increased as if the spins on surface are strongly interacted with the nearest neighbours of core, then it becomes difficult to flip those spins easily. As a consequence, both the surface and bulk reversal times increase. In the higher regime of $R$ an opposite scneraio can be observed where the bulk reversal occurs faster than that of the surface. So clearly, we found the existence of a certain critical relative interaction strength $R_c=(J_f^c/J)$ (the intersection point in the Fig-\ref{3magtime}b) for which metastable lifetimes of the surface and the bulk become almost equal. It is worth mentioning that, by varying $J_f$ or $R$, we get a rough estimation of the effect of the variation of thickness of a sample. In the vanishing limit of $J_f$ or $R$, the surface purely behaves as a two dimensional system.

\noindent $\Box$ {\large \bf Dependence of critical interfacial interaction on temperature and field:}

\begin{figure}[htbp]
	\begin{subfigure}{0.495\textwidth}
	\includegraphics[angle=0,width=\textwidth]{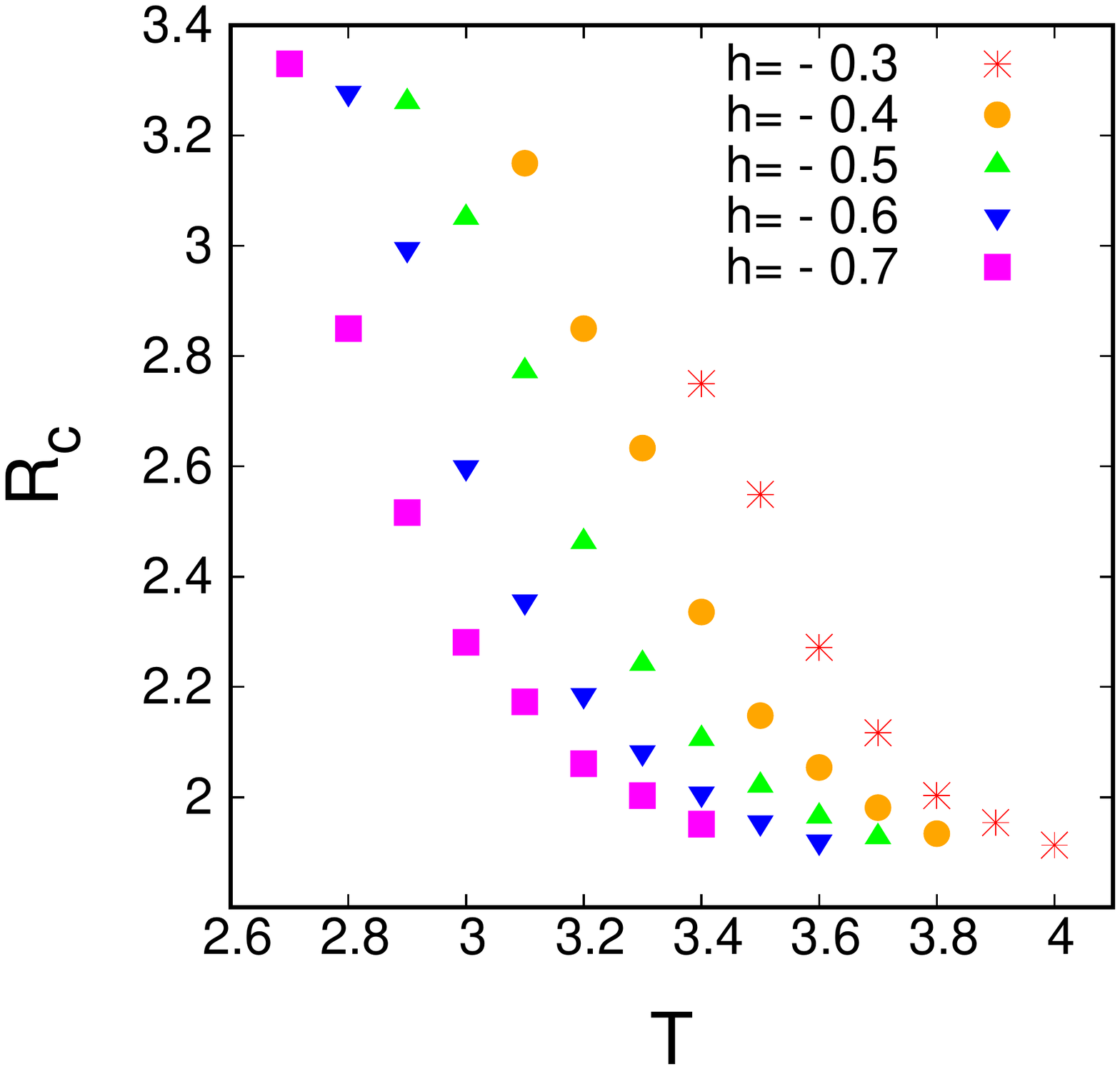}
	\subcaption{}
	{}
	\end{subfigure}
	\begin{subfigure}{0.495\textwidth}
	\includegraphics[angle=0,width=\textwidth]{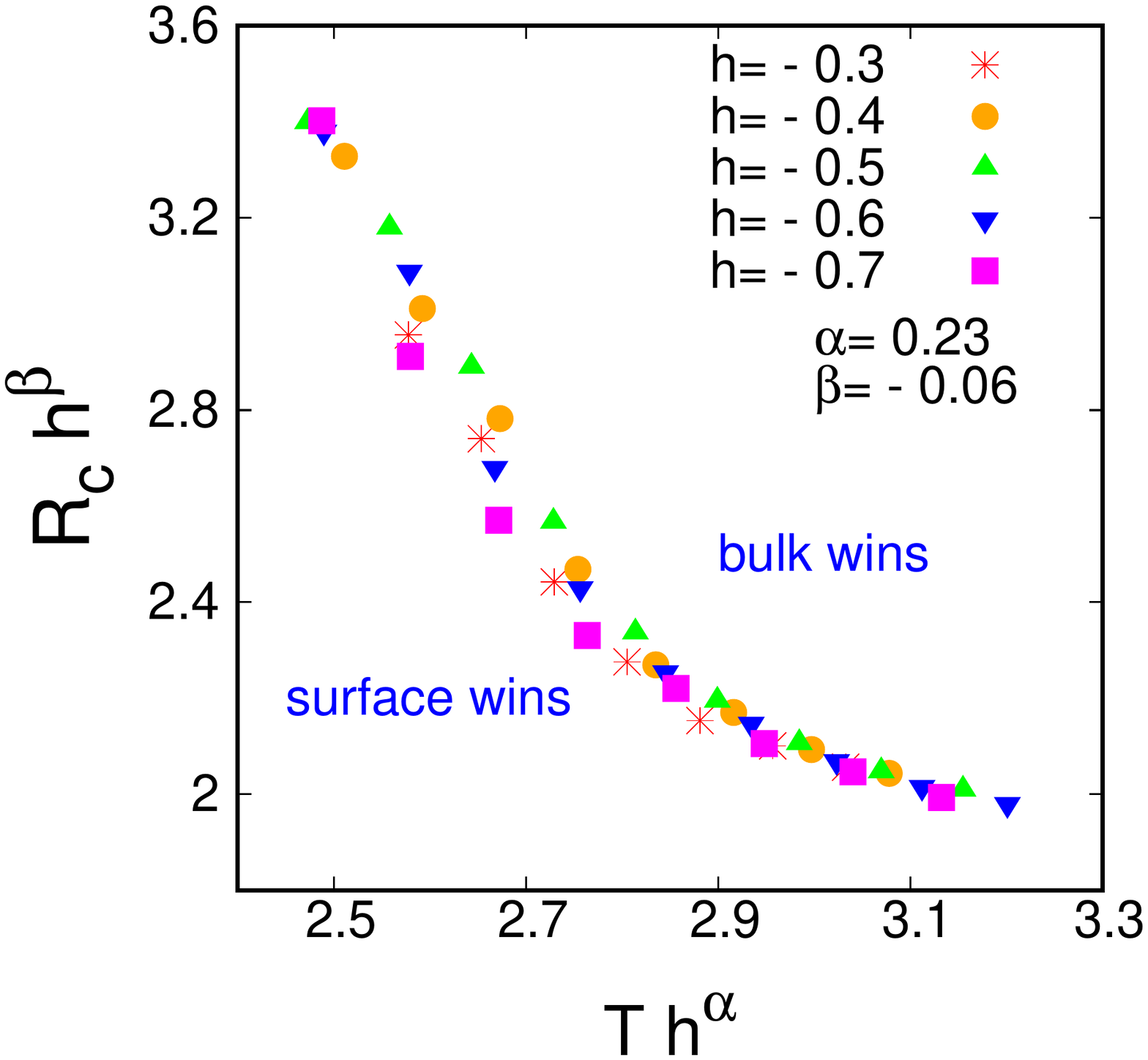}
	\subcaption{}
	\end{subfigure}
\caption{\footnotesize (a) Variation of critical relative interaction strength $R_c$ with temperature $T$ for five different strengths of applied field. (b) Variation of scaled critical relative interaction strength $R_c h^\beta$ with scaled temperature $T h^\alpha$ with exponents $\alpha= 0.23 \pm 0.01$ and $\beta= -0.06 \pm 0.01$. Collapsed data follow the scaling relation $R_c \sim h^{-\beta} f(Th^\alpha)$. The size of the system is $L=32$. Naskar M. and Acharyya M., 2021, Eur. Phys. J. B, {\bf 94}, 140.}
\label{3scaling}
\end{figure}
The dependence of $R_c$ on temperature has been investigated in presence of different strengths of applied field (Fig-\ref{3scaling}a). For a fixed strength of the applied field, the $R_c$ is identified within a suitable range of temperature where the difference of $\tau_s$ and $\tau_b$ is prominent. It may be noted that the radius of convergence (here 0.01) of the two fitted functions is considered as the error of the determination of $R_c$. Errors are not visible here because of its size which is of the order of size of the symbol of the data point. $R_c$ decreases with the increase of temperature and also strength of the applied field ($h$; truly it is the absolute value of field i.e. $|h|$). Interestingly, data are collapsed for scaled relative interaction strength $R_c h^\beta$ and scaled temperature $T h^\alpha$ with $\alpha = 0.23 \pm 0.01$ and $\beta = -0.06 \pm 0.01$. Exponents are optimized (visually) by simple trial and error method to get the data collapsed (Fig-\ref{3scaling}b). So the relative interaction strength ($R$) follows a scaling relation with temperature and applied field, $R_c \sim h^{-\beta}f(Th^\alpha)$. The form of the function $f(Th^\alpha)$ is not yet determined. Now, let me notify an interesting fact that the collapsed data indicates a boundary along which the reversal process of the surface almost synchronizes with that of the bulk. As far as the faster reversal of magnetisation is concerned, below that boundary, the surface-reversal wins over the bulk-reversal. Whereas, above the boundary, the bulk-reversal wins over the surface-reversal.

\begin{figure}[h!]
	\centering
	\includegraphics[angle=0,width=0.5\textwidth]{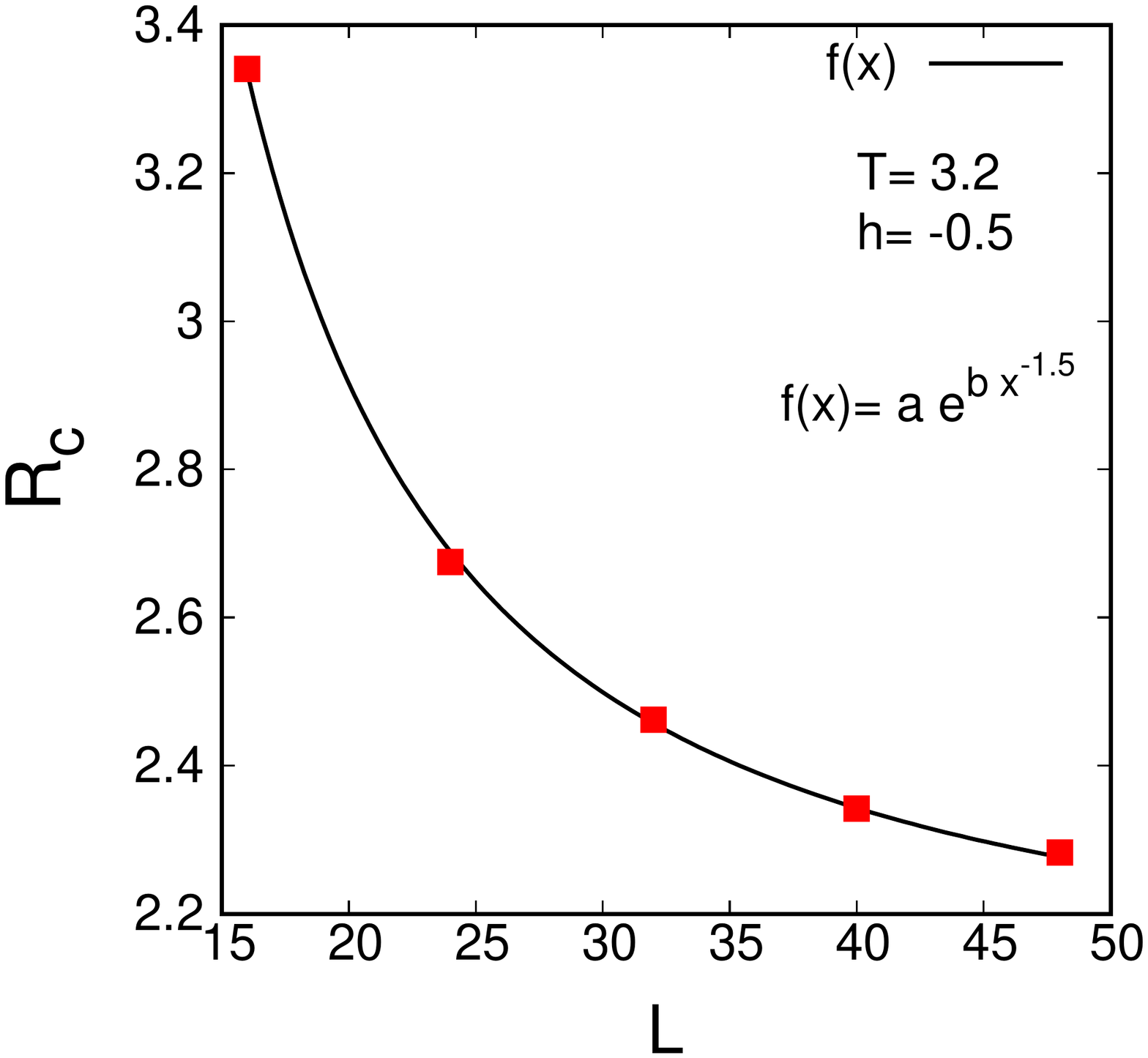}
	\caption{\footnotesize Critical relative interaction strength $R_c$ (for which $\tau_s = \tau_b $) for five different sizes of the system $L=16,24,32,40,48$. Data fit to the function $f(x)= a e^{b x^{-1.5}}$ with $a=2.078 \pm 0.007$ and $b=30.3 \pm 0.3$. Naskar M. and Acharyya M., 2021, Eur. Phys. J. B, {\bf 94}, 140.}
	\label{3fsize2}
\end{figure}

\noindent $\Box$ {\large \bf Finite size effect on critical interfacial interaction}
Mean reversal times of surface $\tau_s$ and bulk $\tau_b$, estimated from 1000 different samples, have been studied with the variation of relative interaction strength $R$ at a fixed temperature $T=3.2$ and applied field $h=-0.5$ for different sizes of lattice $L=16,24,32,40,48$. The dependence of $\tau_s$ and $\tau_b$ on $R$ fit to the exponential function $f(x)= \frac{a}{1+e^{(b-x)/c}}$. $R_c$ is determined for each size of lattice in the same fashion as described before. Radius of convergence (here 0.01) of the two fitted functions is considered as the error (of the order of size of the data point) in order to determine $R_c$. Variation of $R_c \pm 0.01$ with size of the system $L$ fit to the function $f(R_c) \sim a \; e^{(b L^{-1.5})}$ with $a=2.078 \pm 0.007$ and $b=30.3 \pm 0.3$. It is worth mentioning that, in the thermodynamic limit ($L\rightarrow \infty$), the value of $R_c$ tries to reach a fixed value $R_c \simeq 2.078$ (Fig-\ref{3fsize2}).


\vskip 0.5 cm

\subsection{Effects of magnetic anisotropy on reversal of magnetisation}

In the modern technologies of magnetic memory devices, magnetic anisotropy plays a crucial role. Due to magnetic anisotropy spins try to align along a preferred direction often disregarding the direction of an externally applied field which makes the magnetic properties of the system direction-dependent. The spin-1 Blume-Capel model is the simplest choice to study such effects. The phase diagram of an anisotropic system was numerically explored by Blume-Capel (Blume, 1966; Capel, 1966) model by including an extra anisotropic term to the Hamiltonian compared to the Ising system. In the last few years, various behaviours of an anisotropic system have been elucidated using the Blume-Capel model. The thermally activated magnetisation switching of small ferromagnetic particles driven by an external magnetic field has been investigated and interestingly a crossover from coherent rotation to nucleation for a classical anisotropic Heisenberg model has been reported (Hinzke and Nowak, 1998). Metastability and nucleation in the Spin-1 Blume-Capel (BC) ferromagnet was explored and found the different mechanism of transition (Cirillo and Olivieri, 1996). 

\begin{figure}[htbp]
\begin{center}
\begin{subfigure}{0.495\textwidth}
\includegraphics[angle=0,width=1.1\textwidth]{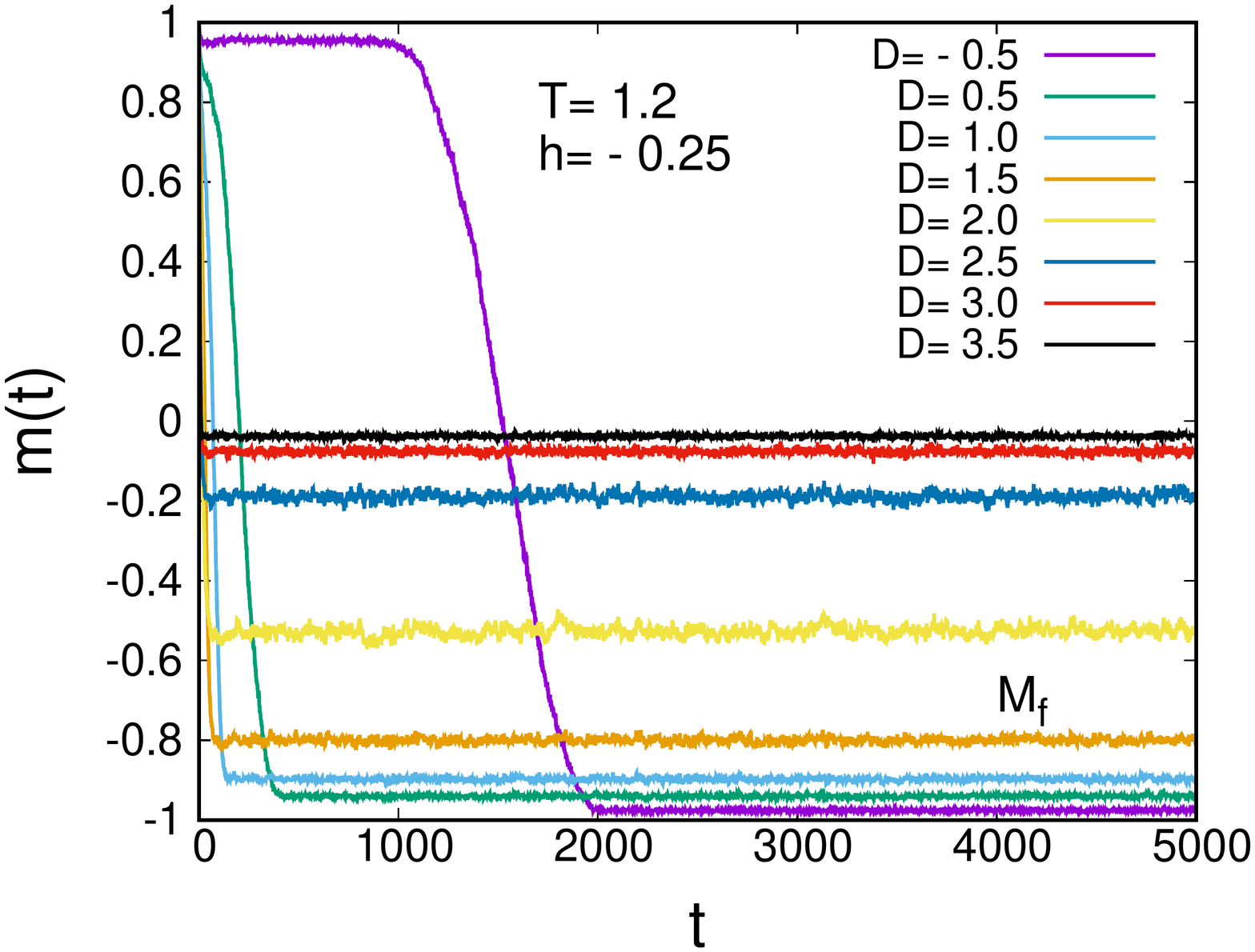}
\subcaption{}
\end{subfigure}

\caption{\footnotesize (a) Variation of magnetisation with time for different values of anisotropy at a temperature $T=1.2$ in presence of applied field $h= -0.25$. Naskar M. and Acharyya M., 2021, Eur. Phys. J. B, {\bf 94}, 36.} 

\label{4magtime}
\end{center}
\end{figure}
How does the anisotropy of a system modulate the metastable lifetime or reversal time of magnetisation? Here we have extensively investigated the role of single-site anisotropy in the reversal mechanism of Blume-Capel ferromagnet (Naskar and Acharyya, 2021a). The spin-1 Blume-Capel ferromagnet is modeled by the following Hamiltonian,
\begin{equation}
     \mathcal{H}= -J\sum_{<i,j>}\sigma_i^z \sigma_j^z + D\sum_{i}(\sigma_i^z)^2 - h\sum_{i}\sigma_i^z
\end{equation}
where $\sigma_i^z$ can assume three values, 1, 0 and -1. Update of this square Blume-Capel lattice with periodic boundary conditions on both directions follow the numerical protocol as described in the section-\ref{model}.


\vskip 0.5 cm

\noindent $\Box$ {\large \bf Metastable lifetime in presence of anisotropy:}
Time evolution of the magnetisation ($m(t)$) 
\noindent of a single sample has been studied for different strengths of anisotropy $D$ at a fixed temperature $T=1.2$ in presence of an externally applied negative field $h=-0.25$. Fig-\ref{4magtime}a depicts that the reversal time $\tau$ of the magnetisation decreases with the increase of the strength of anisotropy ($D>0$). Additionally, we noticed that the saturation magnetisation $M_f$, after the complete reversal, also varies with the strength of anisotropy. $M_f$ is determined by taking the time average of the magnetisation after reaching saturation (flatness of the plots in negative magnetisation region in Fig-\ref{4magtime}a). 
In the case of negative anisotropy ($D=-0.5$), $M_f$ reaches a negative value (close to -1) i.e. a considerably large number of the spins are flipped to $\sigma_i^z=-1$ state (along the direction of the applied magnetic field). In contrast, for positive anisotropy ($D > 0$), $|M_f|$ decreases with increasing the magnitude of the strength of the anisotropy and finally reaches zero. Actually, for negative $D$, the z-axis becomes the easy axis and ultimately the system behaves as a spin-1/2 Ising system in the large limit of negative $D$. 
But for positive anisotropy, z-axis becomes the hard axis and spins favour to access `0' value for minimizing the energy. As a result, the mean density of $\sigma_i^z=0$ starts to grow as the magnitude of the positive anisotropy is increased. It may be notified that, due to large positive anisotropy, the value of the magnetisation of the system is mostly determined by $D$ unlike the situation for negative $D$ where it was preferably determined by the applied magnetic field.

\begin{figure}[htbp]
\centering
	\begin{subfigure}{0.495\textwidth}
	\includegraphics[angle=0,width=1.1\textwidth]{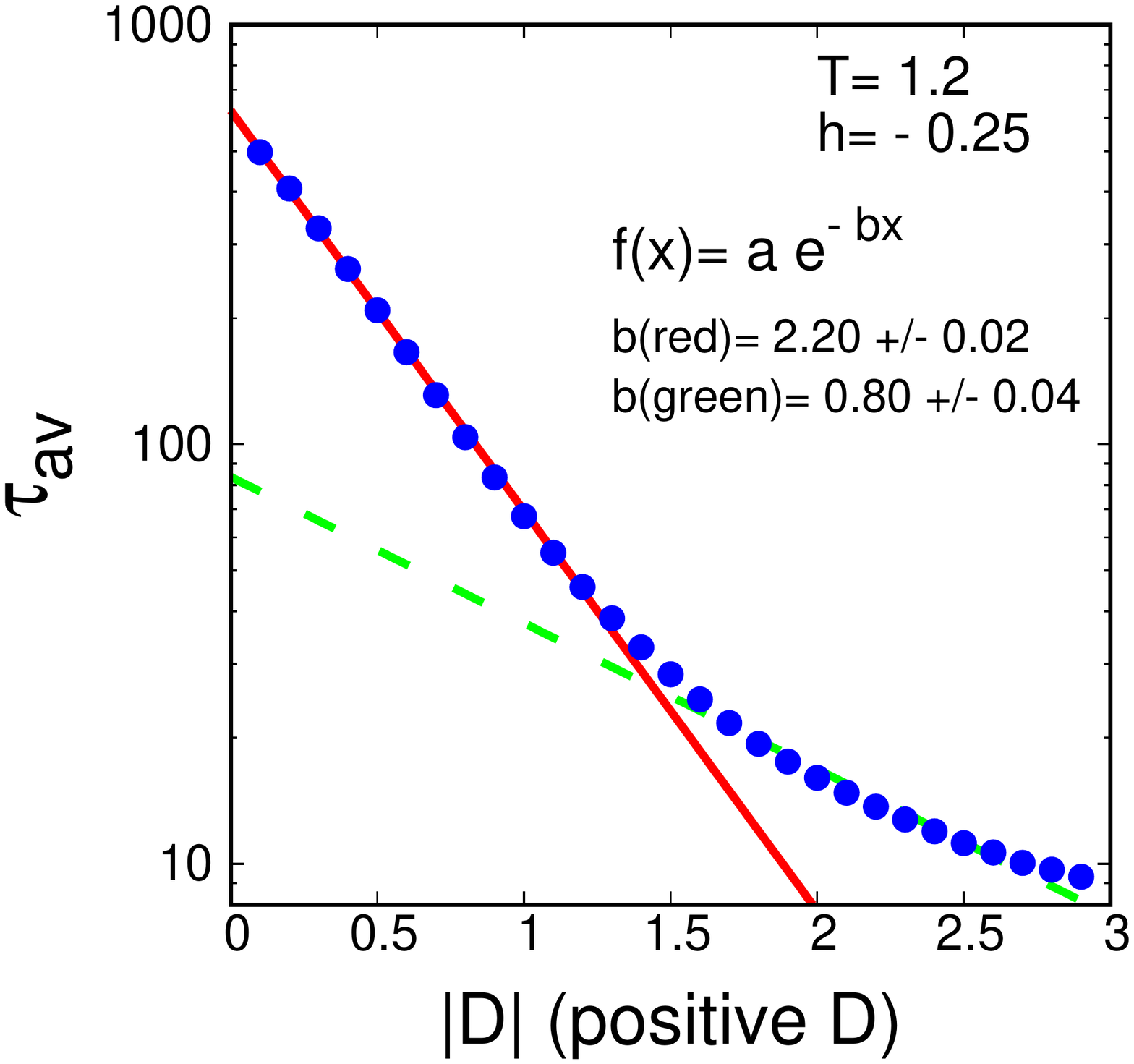}
	\subcaption{}
	\end{subfigure}
	\begin{subfigure}{0.495\textwidth}
	\includegraphics[angle=0,width=1.1\textwidth]{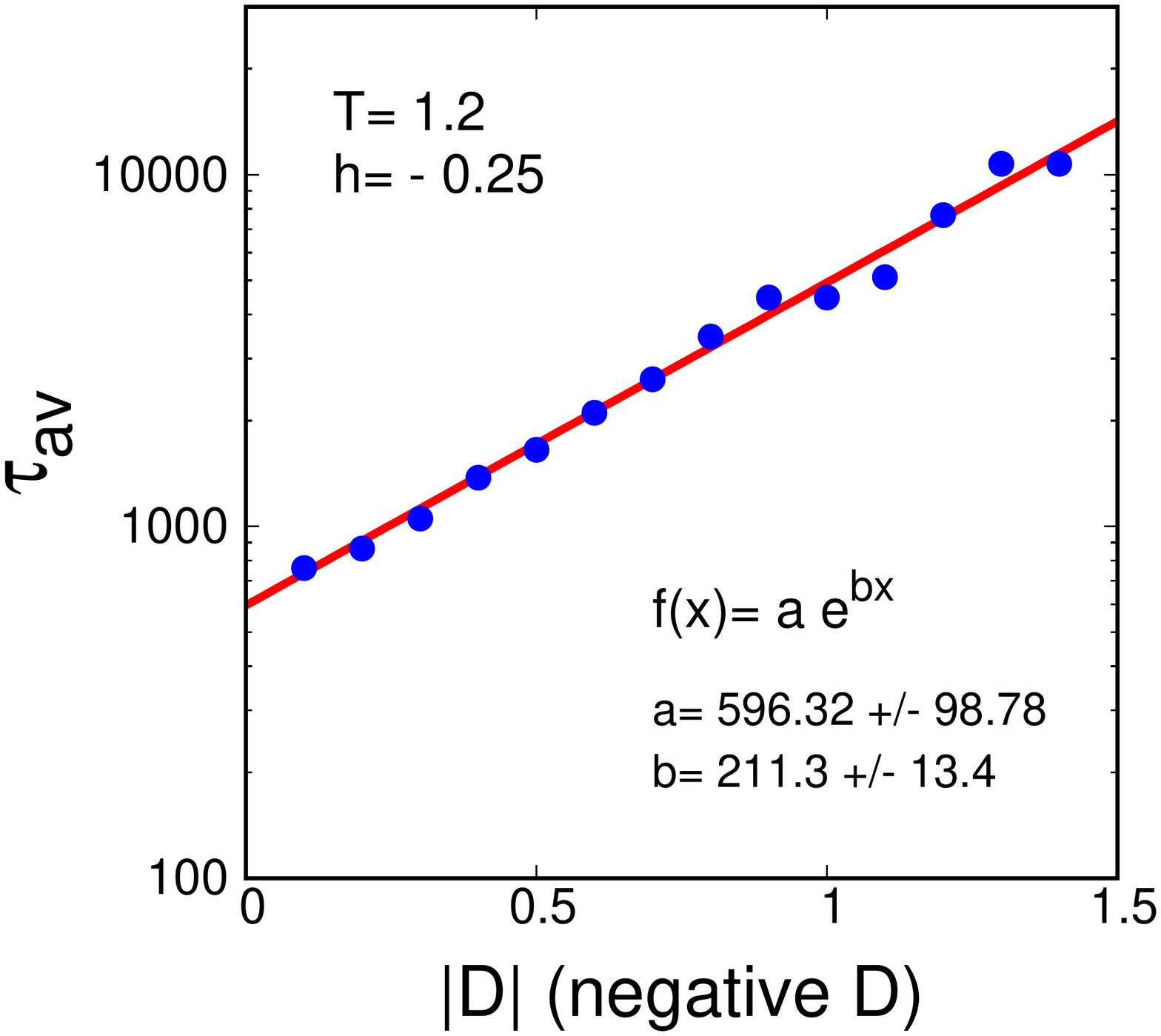}
	\subcaption{}
	\end{subfigure}
\caption{\footnotesize Semilogarithmic plot of variation of mean reversal time $\tau_{av}$, obtained from the 10000 random samples with (a) positive anisotropy and (b) negative anisotropy. Temperature is $T=1.2$ and the applied field is $h= -0.25$. Naskar M. and Acharyya M., 2021, Eur. Phys. J. B, {\bf 94}, 36.}
\label{4revtime_d}
\end{figure}

Variation of the mean reversal time $\tau_{av}$, determined over 10000 sample, with the strength of both positive and negative anisotropy has been checked (Fig-\ref{4revtime_d}). 
Positive $D$ indicates the `z' axis as hard axis whereas negative $D$ confirms the `z' axis as easy axis. The $\tau_{av}$ is found to decrease exponentially ($\tau_{av} \sim e^{-gD}$) with the increase of positive anisotropy and to increase exponentially ($\tau_{av} \sim e^{-g'D}$) with the increase of the absolute value of anisotropy. Let me mention that, if we merge the plots by considering the sign of $D$, then it would be a single plot. We have studied them separately as the sign of $D$ carry different physical significance.   
As already mentioned earlier that, for $D>0$, the z-axis becomes the hard axis and so most of the spins will flip to `0' state. As a result, the absolute value of magnetisation decreases due to the production of a large number of $\sigma_i^z=0$ (which contributes nothing to the magnetisation). On the other hand, a stronger value (magnitude) of negative $D$ will map the system onto an equivalent spin-1/2 Ising ferromagnet, where the single spin-flip would require more cost of energy than that of a Blume-Capel ferromagnet with positive $D$, which has a possibility of transition from $\sigma_i^z=1$ to $\sigma_i^z=0$. This is a possible reason for getting a smaller reversal time in the case of larger positive $D$ in the BC model. For the case of $D>0$, the presence of a crossover has been noticed in the variation of both the $\tau_{av}$ and $\sigma_\tau$. That may be the reflection of the appearance of first-order phase transition influenced by the anisotropy (Butera and Pernici, 2018).

\vskip 0.5cm

\vskip 0.5cm
\noindent $\Box$ {\large \bf Relation between macroscopic and microscopic switching time:}
Macroscopic reversal is definitely connected to some microscopic switching of some spin arrangements. In order to
\begin{figure}[h!]
\centering
\begin{subfigure}{0.495\textwidth}
\includegraphics[angle=0,width=1.1\textwidth]{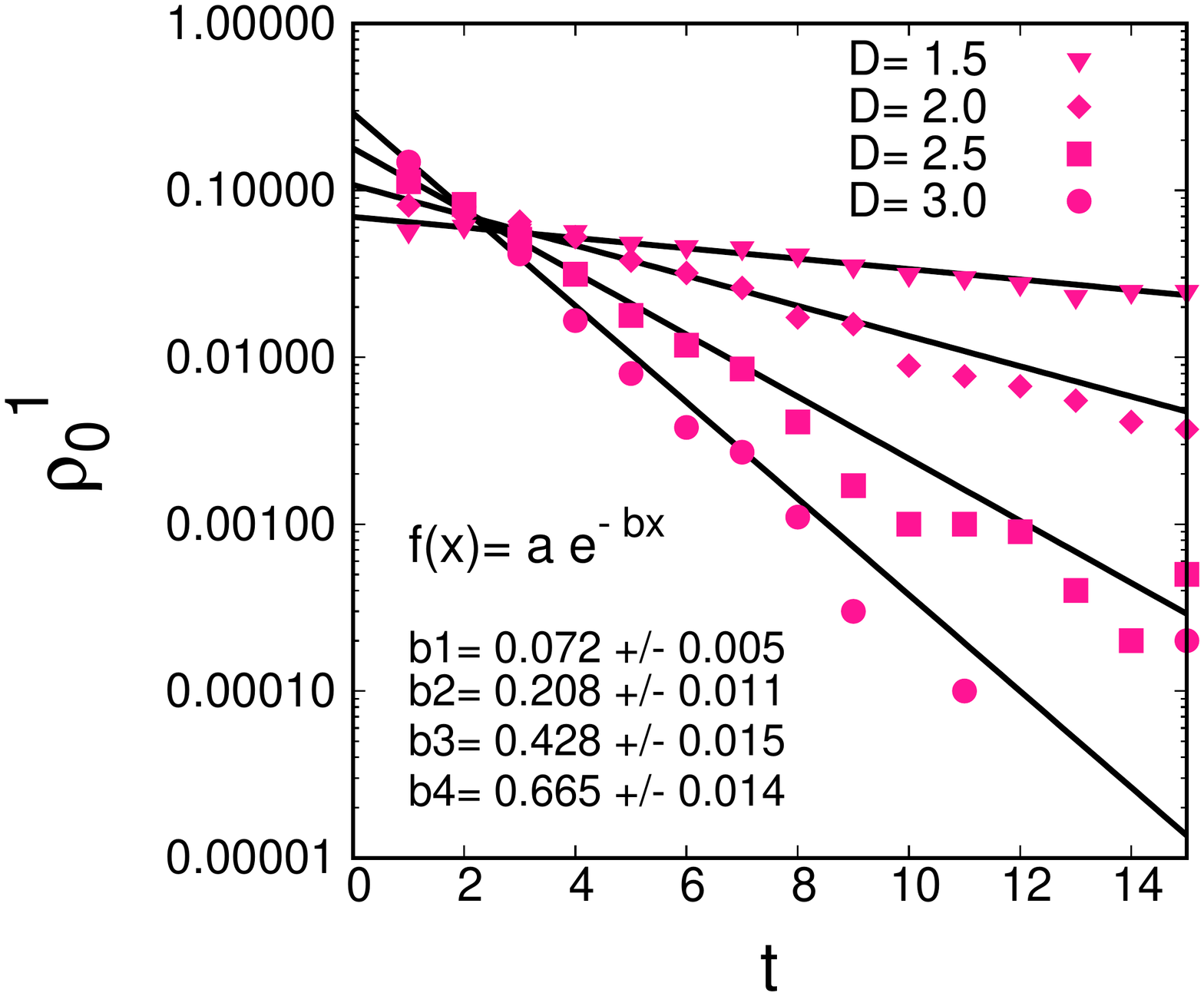} 
\subcaption{}
\end{subfigure}
\begin{subfigure}{0.495\textwidth}
\includegraphics[angle=0,width=1.1\textwidth]{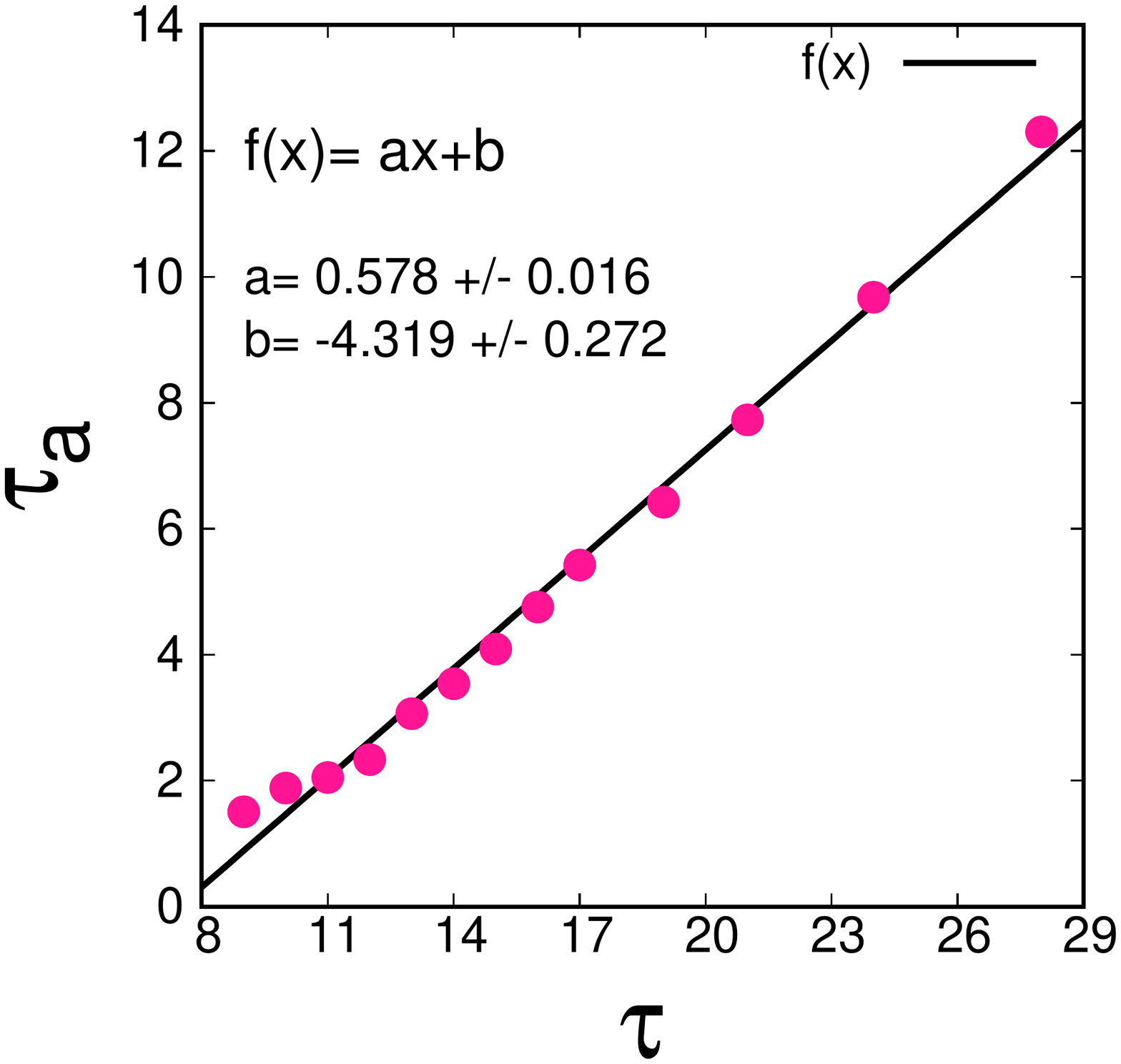}   
\subcaption{}
\end{subfigure}
\caption{\footnotesize (a) Temporal evolution of the density of $\sigma_i^z=0$ surrounded by all neighbouring four $\sigma_i^z=1$ ($\rho_0^1$) for four different strengths of anisotropy (here, $D= 1.5, 2.0, 2.5, 3.0$). Data fit to the function $\rho_0^1= a e^{-bt/10}$. Applied field is $h= -0.25$ and temperature is $T=1.2$. (b) Relation between microscopic reversal time ($\tau_a$) and average macroscopic reversal time $\tau$. The value of $\tau_a = \frac{10}{b}$ is calculated from Fig-\ref{4revtime_confirm1}a. Data fit to a straight line $\tau_a= a \tau +b$. Naskar M. and Acharyya M., 2021, Eur. Phys. J. B, {\bf 94}, 36.}
\label{4revtime_confirm1}
\end{figure}
explore that, we have studied (Fig-\ref{4revtime_confirm1}a) the evolution of density of $\sigma_i^z=0$ ($\rho_0^1$), surrounded by all (four nearest neighbours) $\sigma_i^z=1$, with time in presence of anisotropy ($D>0$ here). The justification of considering this microscopic configuration, is that for the magnetisation reversal for positive values of $D$, such microscopic configuration is dominantly effective to produce more $\sigma_i^z=0$ in the system to yield macroscopic reversal eventually. That density $\rho_0^1$ is found to decay exponentially ($\rho_0^1= a e^{-bt}$) with time. The microscopic scale of time is defined as $\tau_a= \frac{1}{b}$ is considered as the microscopic scale of time in the present issue. Now the $\tau_a$ is determined for several values of anisotropy and plotted with the reversal time ($\tau$) that we have defined earlier. It follows a straight line (Fig-\ref{4revtime_confirm1}b, $\tau_a \sim c \tau$, where $c$ is a constant). This interesting observation prompted us to have the idea of getting the microscopic scale of time ($\tau_a$) which is related to the macroscopic reversal time ($\tau$). It may be noted here, that both time scales are measured in the case of a single sample only (no averaging is carried out over different random samples).

\begin{figure}[htbp]
\centering
	\begin{subfigure}[b]{0.49\textwidth}
	\includegraphics[angle=-90,width=\textwidth]{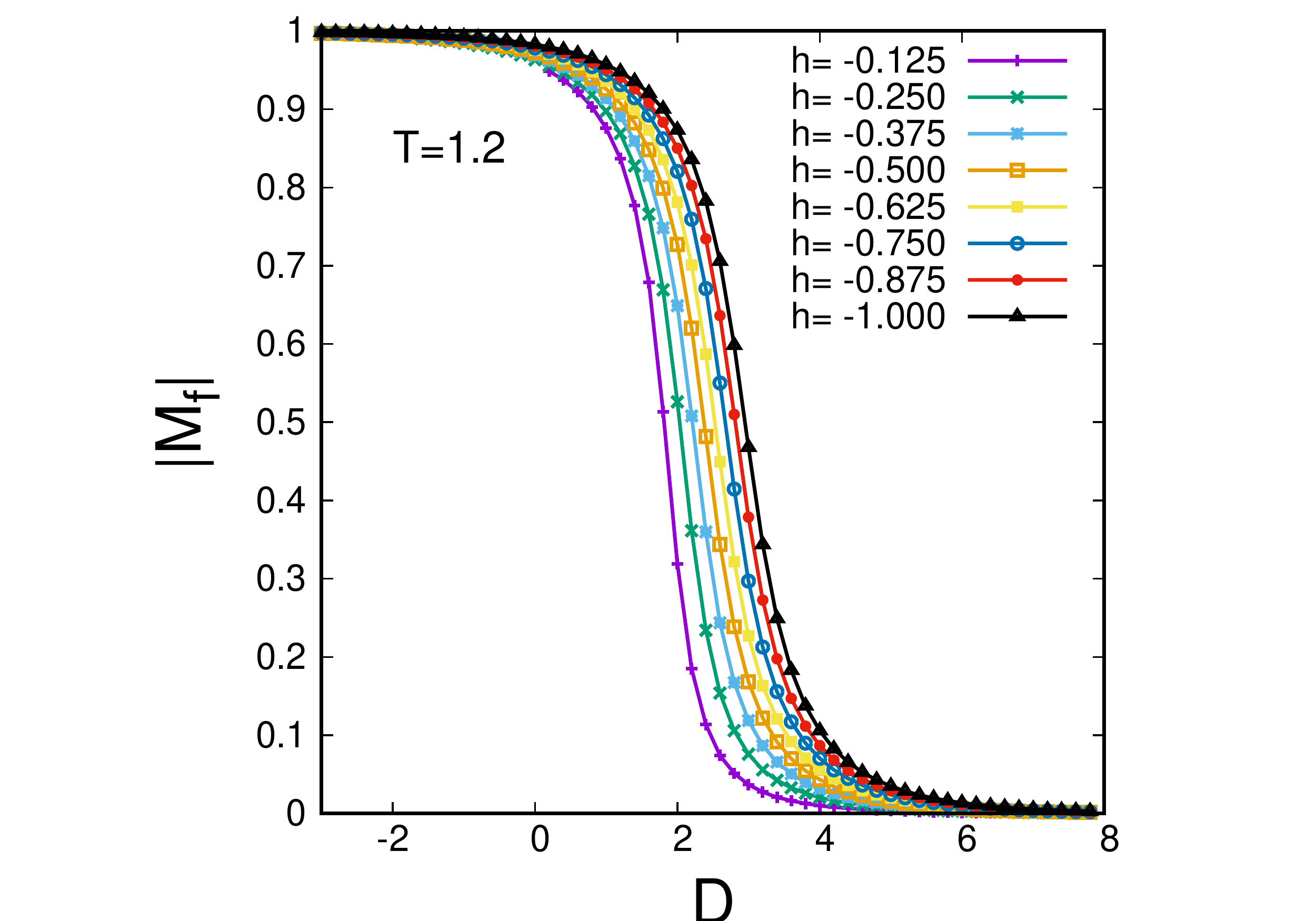}
	\subcaption{}
	\end{subfigure}
	\begin{subfigure}[b]{0.49\textwidth}
	\includegraphics[angle=-90,width=\textwidth]{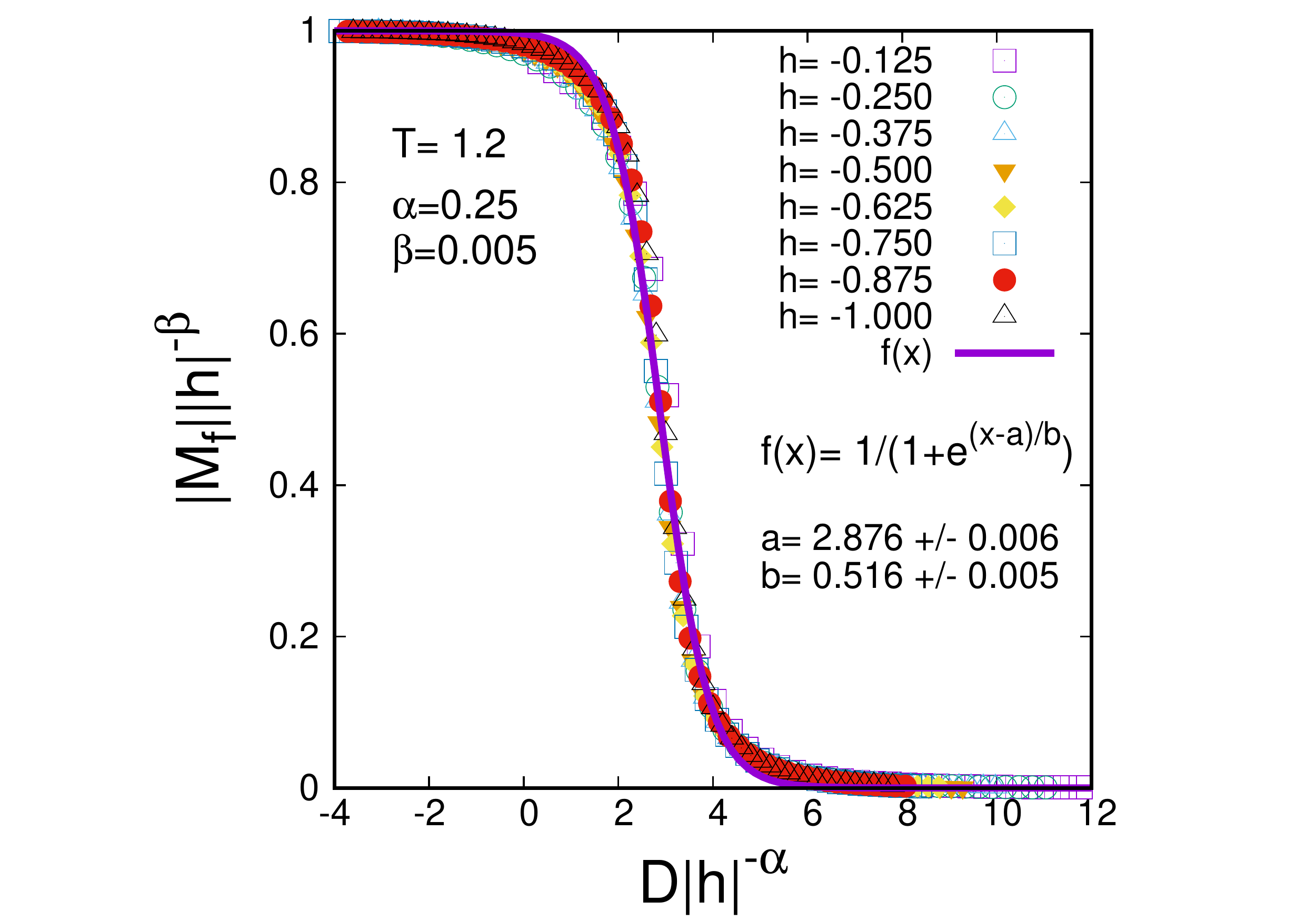}
	\subcaption{}
	\end{subfigure}
\caption{\footnotesize (a) Variation of saturated magnetisation ($M_f$) with anisotropy $(D)$ at any fixed temperature $T=1.2$ for different values of applied field ($h$), (b) Scaled saturated magnetisation ($|M_f||h|^{-\beta}$) versus scaled anisotropy ($D|h|^{-\alpha}$) at fixed temperature $T=1.2$. Naskar M. and Acharyya M., 2021, Eur. Phys. J. B, {\bf 94}, 36.}
\label{4scaling1}
\end{figure}

\vskip 0.5cm
\noindent $\Box$ {\large \bf Effect of anisotropy on the post-reversal saturation magnetisation:}
Now we will discuss the role of anisotropy in the saturation magnetisation ($|M_f|$) after complete reversal. In Fig-\ref{4scaling1}a, the dependence of ($|M_f|$) on the anisotropy of the system is checked in presence of different values of applied field at a fixed temperature $T=1.2$. $|M_f|$ is found to follow a scaling relation $|M_f| \sim |h|^{\beta} D|h|^{-\alpha}$ obtained by using data collapse technique by simple trial and error method where $\alpha=0.25$ and $\beta=0.005$ (Fig-\ref{4scaling1}b). Collapsed data fit to a scaling function, $f(x)= \frac{1}{1+e^{(x-a)/b}}$ where $f(x)= |M_f||h|^{-\beta}$ and $x=D|h|^{-\alpha}$. Additionally, it is observed that the scaling exponent $\alpha$ plays the crucial role ($\beta$ is quite small) in collapsing the data. For $D<0$, all the spins try 
to align along (parallel or antiparallel) the direction of the applied field. That results in perfect magnetic ordering. In contrast, the strong positive anisotropy forbids the spins to be aligned along (parallel or antiparallel) the direction of the applied magnetic field. Rather, spins try to align along the direction perpendicular to the applied field as the `z' axis becomes hard axis. As a consequence, the system seems to exhibit no magnetic ordering ($M_f \simeq 0$). But it is worth mentioning here that, the system actually exhibits different kinds of magnetic ordering which is influenced by the anisotropy and not by the applied magnetic field.

\subsection{Reversal in spin-s Ising and Blume-Capel system}

Now, we will try to extend the study by addressing the important next-step question of whether the predictions of classical nucleation theory can be observed at any general spin-$s$ Ising-type ferromagnet (Naskar et al, 2021). This open problem of understanding reversal processes in magnets with high spin values is not only of great theoretical interest but also strongly connected to the developing modern technologies which are based on controlled switching of the spin state. Recently many investigations are going on depending on the switching of spin states (Shankar et al., 2018, Ohkoshi et al., 2002, Reed and Guiset, 1996). In this regard, some complex ions are purposefully used which have more than one electron in their outermost shell, and coupling between those electrons gives rise to high spin states. To this end, the study in this chapter will attempt to provide some clear answers to the following open questions: Does the Becker-D\"{o}ring analysis hold for the general case of the spin-$s$ Ising model?

Square lattice of size $L = 100$ of both the Ising and Blume-Capel system has been simulated with \textit{periodic boundary conditions} on both sides.
Ideally, we would like to have a rough estimate of the critical temperature $T_{\rm c}$ of the normalised spin-$s$ Ising and Blume-Capel models so that we can check the Becker-D\"{o}ring results by keeping the system well below the critical temperature. For that, we have followed the simplest way by defining the pseudocritical temperature, $T_{L}^{\ast}$ which is determined by detecting the peak value of the magnetic susceptibility. However, the situation is
\begin{figure}[ht!]
\centering
  \begin{subfigure}{0.49\textwidth}
  \centering
  \includegraphics[width=\linewidth]{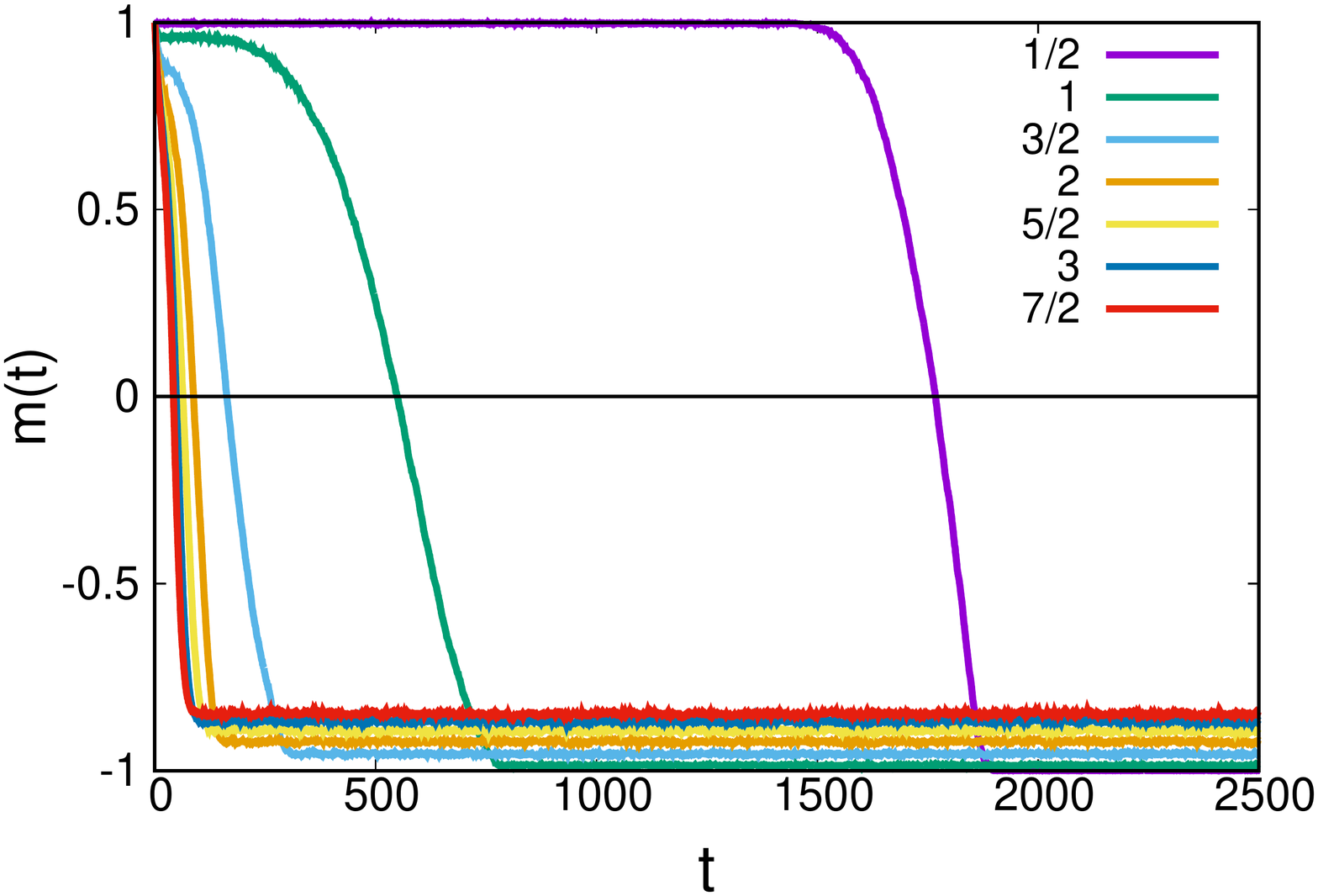}
  \subcaption{}
  \end{subfigure}
  \begin{subfigure}{0.49\textwidth}
  \centering
 \includegraphics[width=\linewidth]{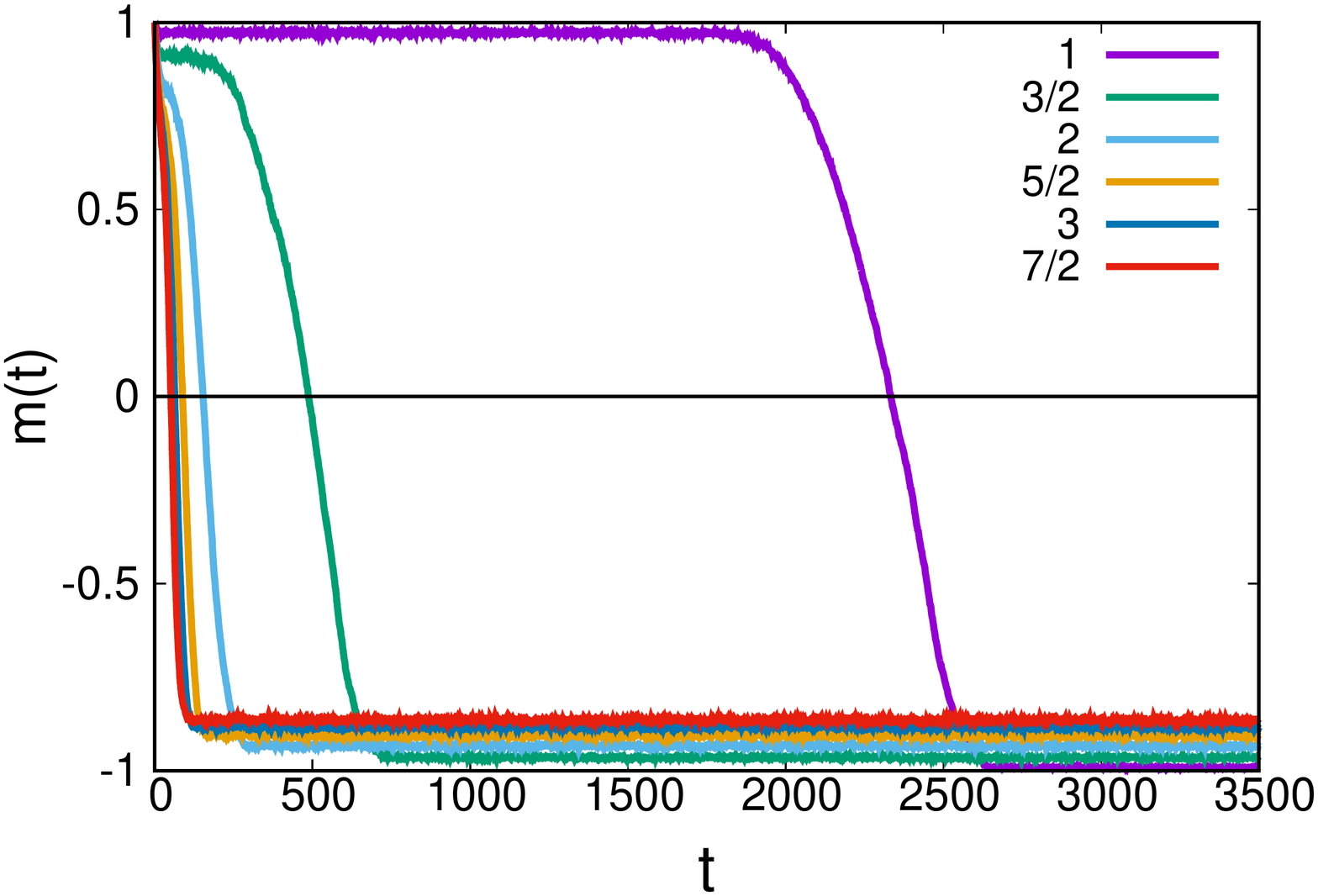}
  \subcaption{}
  \end{subfigure}
\caption{\footnotesize 
. 
Variation of the magnetisation with time (in MCSS) for (a) Ising models at $T=1.0$ and $h = -0.4$ and (b) for BC models having $D=0.5$ at $T=0.8$ and $h = -0.4$. 
Naskar M. and Acharyya M., 2021, Phys. Rev. E, {\bf 104} 014107.} 

\label{6istc}
\end{figure}

In this respect, we have studied the variation of the equilibrium magnetisation $m$ and the corresponding susceptibility $\chi$ for both the Ising and Blume-Capel systems. 
 
\begin{table}[ht!] 
	\caption{\footnotesize Pseudocritical temperatures of the $L = 100$ spin-$s$ Ising and Blume-Capel models obtained from the peak location of the magnetic susceptibility. A uniform maximum error of $10^{-2}$ stems from the temperature-step $\delta T = 0.01$ in our simulations. Naskar M., Acharyya M., 2021, Phys. Rev. E., {\bf 104},
	014107.}
	\centering
	\begin{tabular}{ c|c|c }
		\hline\hline
		Spin-$s$ systems & Ising models & Blume-Capel models\\
		 $s$ & $T_{L}^{\ast}(D=0)$ & $T_{L}^{\ast}(D=0.5)$ \\  \hline	
		1/2 & 2.27(1) & (not applicable) \\		
		1 & 1.72(1) & 1.58(1)\\			
		3/2 & 1.47(1) & 1.33(1)\\		
		2 & 1.34(1) & 1.21(1)\\  		
		5/2 & 1.26(1) & 1.13(1)\\	
		3 & 1.21(1) & 1.08(1)\\			
		7/2 & 1.18(1) & 1.04(1)\\
		\hline \hline		
	\end{tabular}
\label{6tctable}
\end{table}
In Tab.~\ref{6tctable} a summary of approximate pseudocritical temperatures (considered up to second decimal place) is provided for the spin-$s$ Ising and Blume-Capel models ($D=0.5$). The susceptibility is checked by varying the temperature in steps of $\delta T = 0.01$ so that the maximum error associated with the approximate $T_L^{\ast}$ is of the order $\sim 10^{-2}$. In Naskar et al., 2021 we observe that the critical temperature decreases with increasing $s$. In the spin-$1/2$ Ising system, the spin can either access the state `$+1$' or `$-1$'. As the number of spin components increases the system walks through some intermediate accessible states between `$+1$' and `$-1$'. Then the activation energy needed to flip the spin from `$+1$' or `$-1$' via some intermediate state will be much smaller compared to the direct flipping. Furthermore, in the disordered state, the spin-$s$ system will be equally distributed among all of its accessible states. For these reasons, the system consumes lower energy (thermal activation-energy) to be driven into a fully disordered state.

As a result, at a particular temperature, the reversal time is found to decrease with the increasing $s$ in the presence of a uniform magnetic field (\ref{6istc}) in both the Ising and BC model. This is due to the increase in the effective thermal fluctuation (as the value of $s$ increases) which is connected to the fact that as we go to the higher spin values $T_{\rm c}$ decreases.

\vskip 0.5cm

\begin{figure}[h!]
\centering

  \begin{subfigure}{0.495\textwidth}
  \centering
  \includegraphics[width=1.1\linewidth]{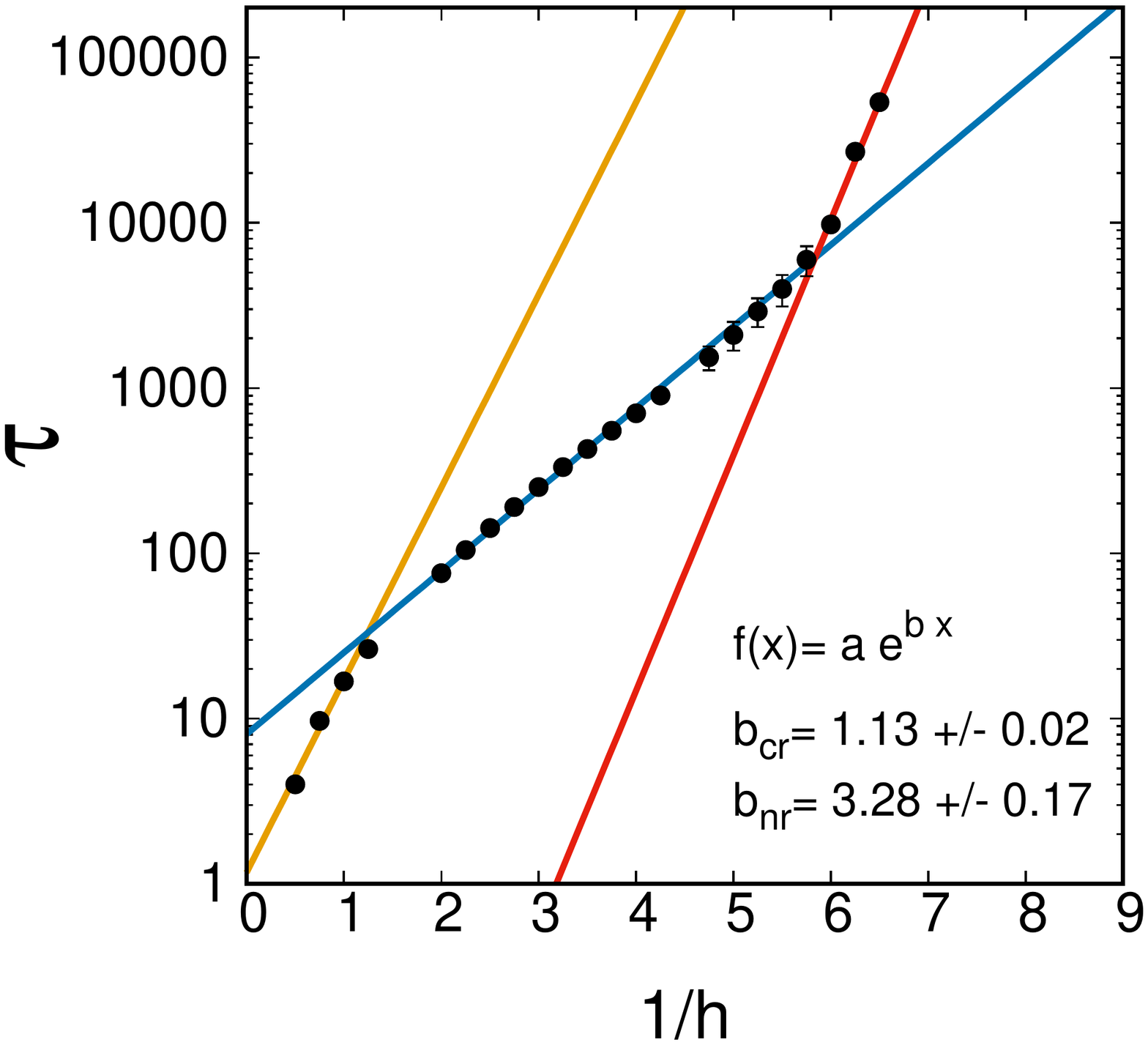}
  \subcaption{$\; s = 2$, $T= 0.94$.}
  \end{subfigure}
  
  \begin{subfigure}{0.495\textwidth}
  \centering
  \includegraphics[width=1.1\linewidth]{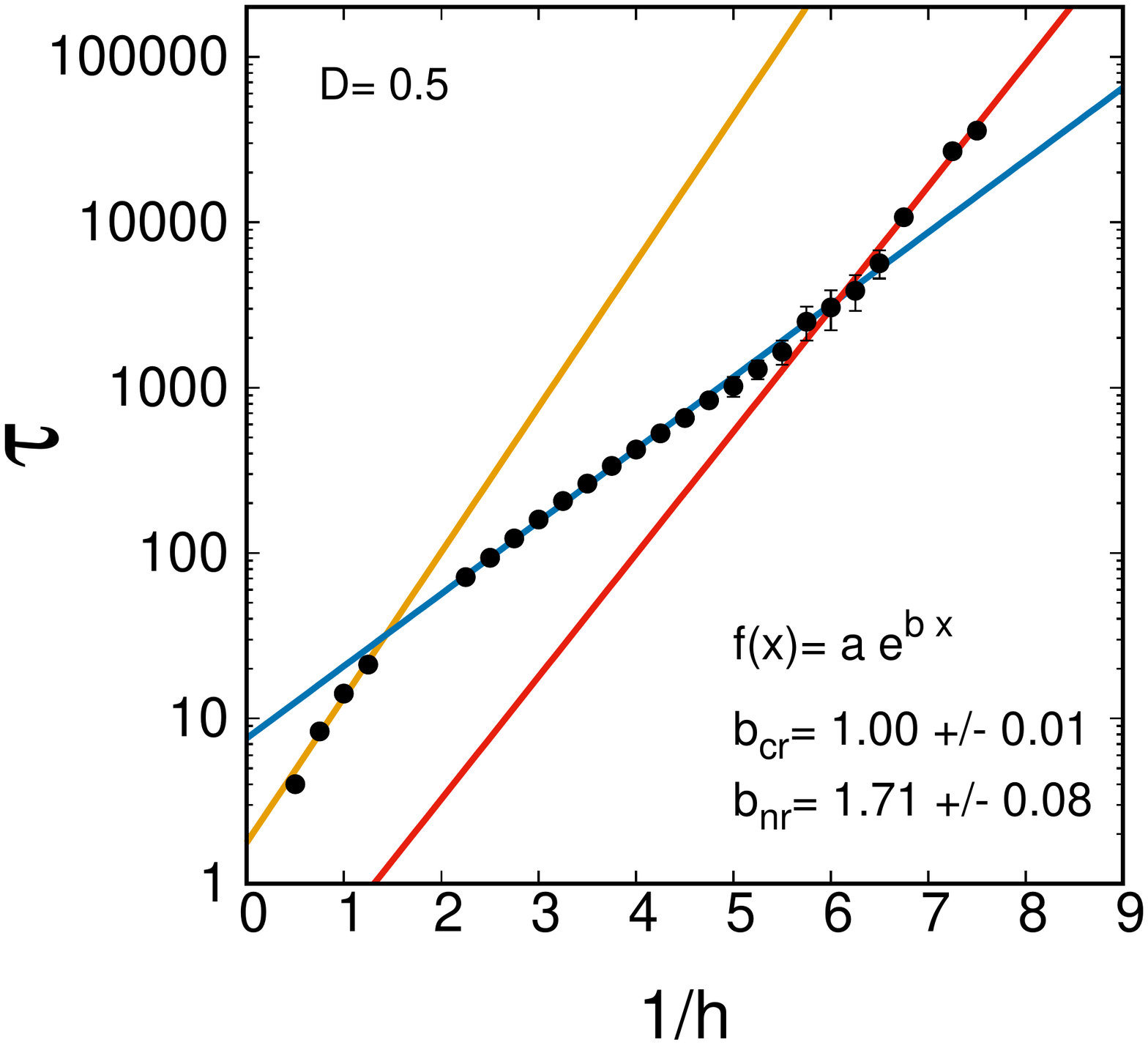}
  \subcaption{$\; s = 5/2$, $T= 0.79$.}
  \end{subfigure}
\caption{\footnotesize Mean reversal time as a function of the inverse magnetic field at $T= 0.7T_{L}^{\ast}$ for four spin-$s$ Ising models, as indicated in the panels. Results averaged over $1000$ samples. In cases where the error bars are not visible, this is due to being smaller than the symbol size used. Note the appearance of three different regimes: (i) strong-field regime (yellow line), (ii) coalescence regime (blue line), and (iii) nucleation regime (red line), which are identified with different slopes. Note the logarithmic scale in the vertical axis.
Naskar M. and Acharyya M., 2021, Phys. Rev. E, {\bf 104} 014107.}
\label{6isingbeck}
\end{figure}

\noindent $\Box$ {\large \bf Verification of Becker-D\"{o}ring prediction on spin-\textit{s} system:}
Fixing now the temperature to $T = 0.7T_{L}^{\ast}$ for each seven system, we checked (Naskar et al, 2021) a variation of the mean reversal time as a function of the inverse magnetic field for the spin-$1/2,1,3/2,2,5/2,3,7/2$ Ising models and the Blume-Capel models at $D=0.5$. Here, we have presented it only for $s=2$ Ising system and $s=5/2$ Blume-Capel system (Fig-\ref{6isingbeck}). Three different regimes with distinct slopes are clearly identified (see also discussion in the figure panels). 
For error estimation, we have used the standard simple block averaging method which involves splitting up the data in small blocks and re-sampling the data by considering the mean of each block. Note that in some cases these error bars are not visible due to being of the order of symbol sizes.

Qualitatively one can clearly argue that the Becker-D\"{o}ring analysis holds good for the general spin-$s$ Ising systems studied. As a side remark let me draw your attention to the fact that, although each system is kept at $T= 0.7T_{L}^{\ast}$, in spin-s Ising system ($s>1/2$) the system enters into the nucleation regime at stronger field compared to the spin-1/2 Ising system. Now if we go back to the droplet theory, the bulk energy term will be modified for the spin-$s > 1/2$ case, since the energy needed to flip the droplet of some intermediate state will be less than the one needed to flip the droplet of spin ``$+1$''. The droplet's formation-energy will then be $E_l=-ahl+\sigma l^{(d-1)/d}$, where $0 < a \le 2$. Obviously, $a=2$ corresponds to the spin-$1/2$ Ising system. Clearly, $h$ need to be large to compensate that decrement in `a'.


\section {Summary:}

Since magnetic storage devices play a crucial role in the data storage of modern civilization, it seems very useful to study the switching time of magnetization in various environments, at least by computer simulation. Here, we have illuminated some of our findings on the basis of Monte-Carlo simulation results using the Ising and Blume-Capel model.

Although the disordered system follows Becker-D\"oring theory of classical nucleation, reversal time is found to decrease in the presence of disorder. Stronger disorder affects the nucleation regime significantly. It drives the system from the nucleation regime to the coalescence regime. The random field disorder plays a similar role as played by temperature. A distinct competitive reversal of the surface and the bulk has been found depending on the strength of an interfacial exchange interaction $R$. A critical value of $R$ is found ($R_C$) for which the reversal of the surface synchronises with that of the bulk. A scaling behaviour is proposed between the $R_C$, temperature and externally applied field. $R_C$ varies with the system size also, which approaches a fixed value in the large scale limit. The reversal time is found to be exponentially dependent on anisotropy ($D$). Macroscopic reversal time is found to be linearly connected to the microscopic switching time. After the reversal, the magnetisation reaches a saturated value (with some fluctuation, of course). This saturated magnetisation $M_f$, follows a scaling relation with anisotropy and field. Avrami’s law holds good even in the anisotropic system.

At a fixed temperature and applied field, reversal time decreases with an increase in the number of spin states. Spin-$s$ Ising and Blume-Capel systems are found to follow Becker-D\"oring theory of classical nucleation which was originally proposed for Ising-1/2 system.

Now, it is worth mentioning a few points regarding the numerical approach followed in the above discussion. Although the Metropolis algorithm may not be the optimum choice for studying the critical properties of the Ising and Blume-Capel models (Bekhechi and Benyoussef, 1997; Plascak et al., 2002), in our studies, it seems to be a safe and convenient vehicle, as we are only interested in the metastable behaviour well below the critical point. Additionally, the main conclusions of our work are expected to be qualitatively insensitive to the use of other algorithms. At a quantitative level, however, one can expect some deviations in the results. For example, it is well-known that local-update and cluster-update algorithms belong to different dynamic universality classes, and so distinct values of metastable lifetimes should be expected.

As future scopes, definitely, some extensions of those works as well as large scale computer simulational results are welcome to enrich the knowledge of metastability in magnetic system. One may simulate other spin models like $q$-state Potts model, clock-model, XY model, anisotropic Heisenberg model etc.

Instead of using the random field, the effect of disorder or impurity can also be modelled by introducing some bond dilution, or site dilution (like random vacancy at some sites) in the system (Berche et al., 2004).

Lastly, the simulational results demand some analytic extension of the Becker-D\"oring theory of classical nucleation so that it can accommodate the effects of disorder, impurity, magnetic anisotropy, etc., although modification of a well-established theory is not an easy task.

\vskip 0.5 cm

\section {Acknowledgements:} MN acknowledges the SVMCM scholarship for financial support. MA thankfully acknowledges the FRPDF Grant from Presidency University, Kolkata. We thank Erol Vatansever and Nikolaos G. Fytas for collaboration.


\vskip 1 cm
\noindent {\bf \Large References}
	
\vskip 0.5cm
Acharyya, M., 2014. Nucleation in Ising ferromagnet by a field spatially spreading
in time, Physica A, 403, 94

Acharyya, M., 2010. Nonequilibrium magnetization reversal by periodic impulsive fields in Ising mean-field dynamics. Physica Scripta 82, 065703.

Acharyya, M., Stauffer, D., 1998. Nucleation and hysteresis in Ising model: classical theory versus computer simulation. Eur. Phys. J. B 5, 571-575.

Becker, R., D\"{o}ring, W., Kinetische Behandlung der Keimbildung in \"ubers\"attigten D\"ampfen. 1935. Ann. Phys. (Leipzig) 416, 719.

Bekhechi, S., Benyoussef, A., 1997. Multicritical behavior of the antiferromagnetic spin-3/2 Blume-Capel model: Finite-size-scaling and Monte Carlo studies. Phys. Rev. B 56, 13954.

Beckmann, B., Nowak, U., Usadel, K.D., 2003. Asymmetric Reversal Modes in Ferromagnetic/Antiferromagnetic Multilayers. Phys. Rev. Lett. 91, 187201.

Berche, P.E., Chatelain, C., Berche, B., Janke, W., 2004. Bond dilution in the 3D Ising model: a Monte Carlo study. Eur. Phys. J. B 38, 463–474.

Binder, K., Heermann, D.W., 1992. Monte Carlo Simulation in Statistical physics, Second edition, Springer-Verlag, Berlin.

Binder, K., M\"uller-Krumbhaar, H., 1974. Investigation of metastable states and nucleation in the kinetic Ising model. Phys. Rev. B 9, 2328.

Blume, M., 1966. Theory of the First-Order Magnetic Phase Change in $UO_2$. Phys. Rev. 141, 517.

Brendel, K., Barkema, G.T., Beijeren, H.V., 2005. Nucleation times in the two-dimensional Ising model. Phys. Rev. E 71, 031601 (2005).

Butera, P., Pernici, M., 2018. The Blume–Capel model for spins $S=1$ and $3/2$ in dimensions $d = 2$ and 3. Physica A 507, 22.

Capel, H.W., 1966. ON THE POSSIBILITY TRANSITIONS OF FIRST-ORDER IN ISING SYSTEMS WITH ZERO-FIELD PHASE OF TRIPLET IONS SPLITTING. Physica (Amsterdam) 32, 966.

Capel, H. W., 1967a, On the possibility of first-order transitions in Ising systems of triplet ions with zero-field splitting II, Physica  33, 295.

Capel, H. W., 1967b, On the possibility of first-order transitions in Ising systems of triplet ions with zero-field splitting II, Physica  37, 423.

Cirillo, E.N.M., Olivieri, E., 1996. Metastability and nucleation for the Blume-Capel model. Different mechanisms of transition. J. Stat. Phys. 83, 473.

Costabile, E., Amazonas, M.A., Viana, J.R. de Sousa, 2012. Study of the first-order transition in the spin-1 Blume–Capel model by using
effective-field theory. Phys. Lett. A 376, 2922.


Deskins, W.R., Brown, G., Thompson, S.H., Rikvold, P.A., 2011. Kinetic Monte Carlo simulations of a model for heat-assisted magnetization reversal in ultrathin films. Phys. Rev. B 84, 094431.

Dhar, A., Acharyya, M., 2016. Reversal of magnetisation in Ising ferromagnet by the field having gradient. Commun. Theor. Phys. 66, 563.

Dutta, R., Acharyya, M., Dhar, A., 2018. Magnetisation reversal in Ising
ferromagnet by thermal and field gradients. Heliyon 4, e00892.

Ellis, M.O.A., Chantrell, R.W., 2015. Switching times of nanoscale FePt: Finite size effects on the linear reversal mechanism. Appl. Phys. Lett. 106, 162407.

Fytas, N.G., Martin-Mayor V., Picco, M., Sourlas, N., Review of recent developments in the random-field Ising model 2018, J. Stat. Phys.
172, 665

Ferrenberg, A.M., Landau, D.P., 1991. Critical behavior of the three-dimensional Ising model: A high-resolution Monte Carlo study. Phys. Rev. B 44, 5081.

Grant, M., Gunton, J.D., 1985. Theory for the nucleation of a crystalline droplet from the melt. Phys. Rev. B 32, 7299.

Gulpinar, G., Vatansever, E., Agartioglu, M., 2012. Effective-field theory with the differential operator technique for a kinetic Blume–Capel model with random diluted single-ion anisotropy. Physica A 391, 3574.

Gunton, J.D., Droz, M., 1983. Introduction to theory of Metastable and Unstable states, springer-verlag, Berlin.

Hinzke, D., Nowak, U., 1998. Magnetization switching in a Heisenberg model for small ferromagnetic particles. Phys. Rev. B 58, 265.

Hinzke, D., Nowak, U., 2002. Simulation of Magnetization Switching in Nanoparticle Systems. phys. stat. sol. (a) 189, 475 (2002). 

Imry, Y., Ma, S.-K., 1975. Random-Field Instability of the Ordered State of Continuous Symmetry. Phys. Rev. Lett. 35, 1399.

Kolesik, M., Richards, H.L., Novotny, M.A., Rikvold, P.A., Lindgard, P.A., 1997. Magnetization switching in nanoscale ferromagnetic grains: Simulations with heterogeneous nucleation. J. Appl. Phys. 81, 5600.

Masrour, R., Jaber, A.,  Magnetic properties of bilayer graphene armchair nanoribbons: A Monte Carlo Study, 2017, J. Magn. Magn. Mater. 426, 225

Naskar, M., Acharyya, M., 2020. Effects of random fields on the reversal of magnetisation of Ising ferromagnet. Physica A 551, 124583.

Naskar, M., Acharyya, M., 2021a. Anisotropy-driven reversal of magnetisation in Blume–Capel ferromagnet: a Monte Carlo study. Eur. Phys. J. B 94, 36.

Naskar, M., Acharyya, M., 2021b. Competitive metastable behaviours of surface and bulk in Ising ferromagnet. Eur. Phys. J. B 94, 140.

Naskar, M., Acharyya, M., Vatansever, E., Fytas, N.G., 2021. Metastable behavior of the spin-s Ising and Blume-Capel ferromagnets: A Monte Carlo study. Phys. Rev. E 104, 014107.

Nowak, U., Heimel, J., Kleinefeld, T., Weller, D., 1997. Domain dynamics of magnetic films with perpendicular anisotropy. Phys. Rev. B 56, 8143.

Ohkoshi, S., Tokoro, H., Utsunomiya, M., Mizuno, M., Abe, M., Hashimoto, K., 2002. Observation of Spin Transition in an Octahedrally Coordinated Manganese(II) Compound. J. Phys. Chem. B 106, 10.

Park, H., Pleimling, M., 2012. Surface Criticality at a Dynamic Phase Transition. Phys. Rev. Lett. 109, 175703.

Piramanayagam, S.N., Chong, T.C., Development in Data Storage: Material Perspective, 2011. Wiley-IEEE Press.

Plascak, J.A., Landau, D.P., 2003. Universality and double critical end points. Phys. Rev. E 67, 015103(R).

Plascak, J.A., Moreira, J.G., Barreto, F.C. s\'{a}, 1993. Mean field solution of the general spin Blume-Capel model. Phys. Lett. A 173, 360.

Puri, S., 1999. Kinetics of phase ordering. Current Science 77, 376.

Reed, C.A., and Guiset, F., 1996. A ``Magnetochemica'' Series. Ligand Field Strengths of Weakly Binding Anions Deduced from $S = 3/2, 5/2$ Spin State Mixing in Iron(III) Porphyrins. J. Am. Chem. Soc. 118, 3281-3282.

Riego, P., Berger, A., 2015. Nonuniversal surface behavior of dynamic phase transitions. Phys. Rev. E 91, 062141.

Rikvold, P.A, Tomita, H., Miyashita, S., Sides, S.W., 1994. Metastable lifetimes in a kinetic Ising model: Dependence on field and system size. Phys. Rev. E 49, 5080.

Schiefele, B., Voivod, I.S., Bowles, R.K., Poole, P.H., 2013. Heterogeneous nucleation in the low-barrier regime. Phys. Rev. E 87, 042407.

Selke, W., Oitmaa, J., 2010. Monte Carlo study of mixed-spin $S = (1/2, 1)$ Ising ferrimagnets. J. Phys.: Condens. Matter 22, 076004.

Shankar, S., Peters, M., Steinborn, K., Krahwinkel, B., S\"onnichsen, Frank D., Grote, D., Sander, W., Lohmiller, T., R\"udiger, O., Herges, R., 2018. Light-controlled switching of the spin state of iron(III). Nat. Commun. 9, 4750.

Silva, C.J., Caparica, A.A., Plascak, J.A., 2006. Wang-Landau Monte Carlo simulation of the Blume-Capel model. Phys. Rev. E 73, 036702.

Tauscher, K., Pleimling, M., 2014. Surface phase diagram of the three-dimensional kinetic Ising model in an oscillating magnetic field. Phys. Rev. E 89, 022121.

Vatansever, E., Vatansever, Z.Demir., Theodorakis, P.E., Fytas, N.G., 2020. Ising universality in the two-dimensional Blume-Capel model with quenched random crystal field. Phys. Rev. E 102, 062138.

Vehkam\"{a}ki, H., 2006. Classical Nucleation Theory in Multicomponent Systems, Springer.

Vehkam\"{a}ki, H., Ford, I.J., 1999. Nucleation theorems applied to the Ising model. Phys. Rev. E 59, 6483.

Vogel, J., Moritz, J., Fruchart, O., 2006. Nucleation of magnetisation reversal, from nanoparticles to bulk materials. Comptes Rendus Physique 7, 977-987.

Yamamoto, Y., Park, K., 2013. Metastability for the Blume-Capel model with distribution of magnetic anisotropy using different dynamics. Phys. Rev. E 88, 012110.

Yeomans, J.M., Fisher, M.E., 1981. Three-component model and tricritical points: A renormalization-group study. II. General dimensions and the three-phase monohedron. Phys. Rev. B, 24, 2825.

\end{document}